\documentclass[smallextended]{svjour3}     

\usepackage{graphicx}                    
\usepackage{courier}                    
\usepackage{color}                       
\usepackage{breakurl}                         
\usepackage{aas_macros}
\usepackage{amsmath}
\usepackage{amssymb}
\usepackage{hyperref}

\begin{document}

\title{Ground Calibration of Solar X-ray Monitor On-board Chandrayaan-2 Orbiter}

\author{N. P. S. Mithun \href{https://orcid.org/0000-0003-3431-6110}{\includegraphics[scale=1.0]{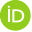}} \and
        Santosh V. Vadawale \href{https://orcid.org/0000-0002-2050-0913}{\includegraphics[scale=1.0]{orcid.pdf}} \and
        M. Shanmugam \and
        Arpit R. Patel \href{https://orcid.org/0000-0002-0929-1401}{\includegraphics[scale=1.0]{orcid.pdf}} \and
        Neeraj Kumar Tiwari \and
        Hiteshkumar L. Adalja \and 
        Shiv Kumar Goyal \href{https://orcid.org/0000-0002-3153-537X}{\includegraphics[scale=1.0]{orcid.pdf}} \and
        Tinkal Ladiya \and 
        Nishant Singh \and 
        Sushil Kumar \and 
        Manoj K. Tiwari \href{https://orcid.org/0000-0001-5143-1423}{\includegraphics[scale=1.0]{orcid.pdf}} \and 
        M. H. Modi \and
        Biswajit Mondal \and
        Aveek Sarkar \href{https://orcid.org/0000-0002-4781-5798}{\includegraphics[scale=1.0]{orcid.pdf}} \and
        Bhuwan Joshi \href{https://orcid.org/0000-0001-5042-2170}{\includegraphics[scale=1.0]{orcid.pdf}} \and
        P. Janardhan \href{https://orcid.org/0000-0003-2504-2576}{\includegraphics[scale=1.0]{orcid.pdf}} \and
        Anil Bhardwaj \href{https://orcid.org/0000-0003-1693-453X}{\includegraphics[scale=1.0]{orcid.pdf}}        
}

\institute{N. P. S. Mithun\\ 
           \email{mithun@prl.res.in}\\\\ 
           N. P. S. Mithun \and Santosh V. Vadawale \and M. Shanmugam \and Arpit R. Patel
           \and Neeraj Kumar Tiwari \and Hitesh Kumar L. Adalja \and Shiv Kumar Goyal \and
           Tinkal Ladiya \and Nishant Singh \and Sushil Kumar \and Biswajit Mondal \and
           Aveek Sarkar \and Bhuwan Joshi \and P. Janardhan \and Anil Bhardwaj \at
           Physical Research Laboratory, Ahmedabad, Gujarat, India 
           \and
           Manoj K. Tiwari \and M. H. Modi           
           \at
           Raja Ramanna Centre for Advanced Technology, Indore, Madhya Pradesh, India        
}

\authorrunning{N. P. S. Mithun et al.} 

\date{Received: date / Accepted: date}

\maketitle


\begin{abstract}
Chandrayaan-2, the second Indian mission to the Moon, carries a
spectrometer called the \textit{Solar X-ray Monitor} (XSM) to perform soft X-ray
spectral measurements of the Sun while a companion payload measures the
fluorescence emission from the Moon. 
Together these two payloads will provide quantitative estimates of
elemental abundances on the lunar surface. XSM is also expected to provide
significant contributions to the solar X-ray studies with its highest
time cadence and energy resolution spectral measurements.
For this purpose, the XSM employs a Silicon Drift Detector and 
carries out energy measurements of incident 
photons in the 1 -- 15 keV range with a resolution of $<$180 eV at 5.9 keV, over
a wide range of solar X-ray intensities.
Extensive ground calibration experiments have been carried out with the XSM
using laboratory X-ray sources as well as X-ray beam-line facilities to
determine the instrument response matrix parameters required for quantitative
spectral analysis. This includes measurements of gain, spectral
redistribution function, and effective area, under various observing
conditions. The capability of the XSM to maintain its spectral performance
at high incident flux as well as the dead-time and pile-up characteristics have 
also been investigated. The results of these ground calibration 
experiments of the XSM payload are presented in this article. 
\end{abstract}

%


\section {Introduction}

The Chandrayaan-2 mission~\cite{vanitha20} includes a remote
X-ray fluorescence spectroscopy experiment to obtain
quantitative estimates of the elemental abundances on the lunar surface.
This experiment has been realized with two instruments onboard the
Chandrayaan-2 orbiter viz. (i) the \textit{Chandrayaan-2 Large Area Soft X-ray Spectrometer}
(CLASS)~\cite{radhakrishna20}, which measures the X-ray fluorescence from the elements
on the lunar surface; and (ii) the \textit{Solar X-ray Monitor} (XSM)~\cite{shanmugam20}, which measures
the spectrum of solar X-rays responsible for excitation of the elements 
on the lunar surface.
From the strength of elemental lines obtained from the CLASS spectrum together with the
incident solar spectrum inferred from the XSM,
the quantitative estimates of abundances of constituent elements on the lunar surface 
can be derived. 
Apart from this, with unique characteristics such as the highest
spectral resolution for a solar broadband soft X-ray spectrometer and the
full resolution spectral measurement at every second, the XSM is also
expected to provide significant contributions to enhance our understanding of 
the corona, the outer atmosphere of the Sun.

The XSM is designed to carry out spectroscopy of the Sun in the 1--15 keV energy range, 
and the basic measurement from the instrument is the number of events registered in each instrument 
channel in every one-second time interval.
However, the elemental analysis as well as any independent investigations
of the Sun would require estimation of the incident solar spectrum,
in physical units, from this observed raw spectrum, which requires
knowledge of various factors contributing to the instrument response.
Uncertainties in the estimates of physical parameters of the source
from these observations would critically depend on how well the 
instrument response has been calibrated. In the case of a space-based experiment like XSM,
the calibration activities involve both the ground calibration where specific
experiments are carried out to determine all the instrument parameters
and the in-flight calibration where these parameters are validated
and further refined.

This article presents an overview of the Chandrayaan-2 XSM instrument and
the results of its ground calibration.
Section~\ref{xsm_inst} describes
the instrument and the calibration requirements are discussed in section~\ref{xsm_gcalreq}.
Sections~\ref{gaincalsec} - \ref{deadtimesec} describe the calibration experiments and their results
followed by a summary in section~\ref{summary}.

\section{XSM on-board Chandrayaan-2}
\label{xsm_inst}

\begin{figure}
 \centerline{\includegraphics[width=0.95\textwidth]{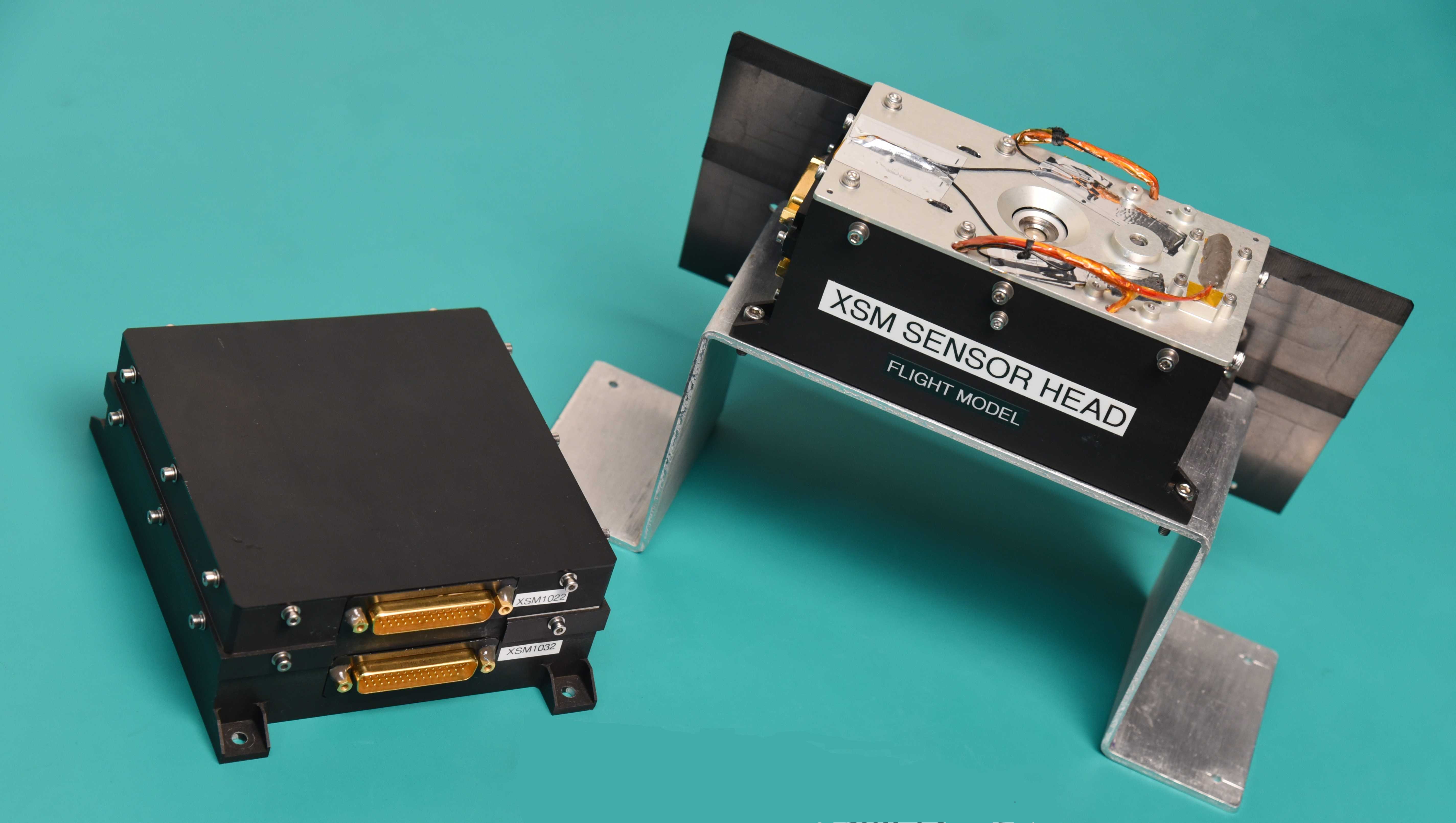}}
 \caption{A photograph of the Chandrayaan-2 XSM instrument packages: 
 (i) sensor package that houses the detector, front-end electronics, and filter wheel mechanism (right) ; 
 (ii) processing electronics package that houses the FPGA based data acquisition system, power electronics, and
 spacecraft interfaces (left).   
}\label{xsm_package}
 \end{figure}

\begin{figure}
 \centerline{\includegraphics[width=0.95\textwidth]{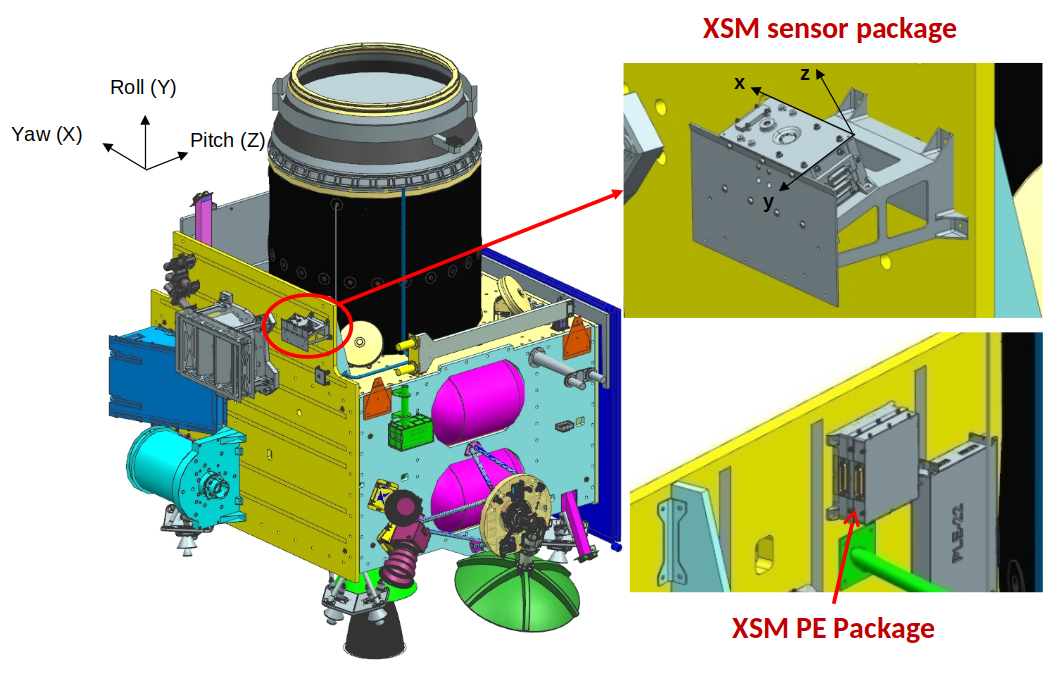}}
 \caption{Schematic representation of Chandrayaan-2 orbiter showing the mounting location 
  of the XSM sensor and processing electronics (PE) packages. 
    The spacecraft reference frame axes, Yaw(X), Roll(Y), and Pitch(Z), are marked in the figure. The sensor package is mounted 
  on the -pitch panel of the spacecraft with a canted bracket ($20^{\circ}$ from roll axis towards -pitch direction) 
  to avoid any spacecraft structures obstructing its field of view. The PE package is mounted on the inner side of 
  the -pitch panel.  Axis definitions for the XSM instrument reference frame are marked on the sensor package. 
  The XSM reference frame is obtained by a rotation of $-110^{\circ}$ about the X-axis of the spacecraft frame.}
\label{xsm_mounting}
 \end{figure}

The XSM carries out spectral measurements of the Sun in soft X-rays 
with an energy resolution better than 180 eV at 5.9 keV.
Major specifications of the XSM instrument are given in table~\ref{xsm_spec}, and 
a detailed description of the instrument design can be seen in Shanmugam \textit{et al.}, 2020~\cite{shanmugam20}.
 Figure~\ref{xsm_package} shows a photograph of the XSM instrument which
 is configured as two packages: (i) a sensor package that houses the detector,
 front-end electronics, and a filter wheel mechanism; (ii) a processing electronics package
 that houses the FPGA based data acquisition system, power electronics, and 
 spacecraft interfaces. The instrument
 is fix-mounted on the Chandrayaan-2 orbiter such that the spacecraft structures do not obstruct the instrument 
 field of view (FOV). 
 Figure~\ref{xsm_mounting} shows the mounting of XSM packages on the spacecraft and definitions of 
 the spacecraft and instrument reference frames.

 \begin{table}
 \caption{Specifications of XSM}
 \label{xsm_spec}
 \begin{tabular}{l l}     
 \hline
 Parameter & Specification \\
 \hline
 Energy Range & 1 -- 15 keV (up to $\sim$80,000 $counts~s^{-1}$) \\
              & 2 -- 15 keV (above $\sim$80,000 $counts~s^{-1}$)\\
 Energy Resolution & $<$ 180 eV @ 5.9 keV \\
 Time cadence & 1 second \\
 Aperture area & 0.367 ${mm}^2$\\
 Field of view    & $\pm$40 degree \\
 Filter wheel mechanism properties & \\
 \hspace{0.5 cm}Positions & 3 (Open, Be-filter, Cal)\\
 \hspace{0.5 cm}Filter wheel movement modes & Automated and Manual \\ 
 \hspace{0.5 cm}Be-filter thickness & $250~\mu m$ \\
 \hspace{0.5 cm}Automated Be-filter movement threshold & 80,000 $counts~s^{-1}$ \\
 \hspace{0.5 cm}Calibration source & Fe-55 with Ti foil \\  
 Detector properties & \\
 \hspace{0.5 cm}Type & Silicon Drift Detector (SDD)\\
 \hspace{0.5 cm}Area & 30 ${mm}^2$\\ 
 \hspace{0.5 cm}Thickness & 450$~\mu m$ \\
 \hspace{0.5 cm}Entrance Window & 8$\mu m$ thick Be \\  
 \hspace{0.5 cm}Operating temperature & -35$^{\circ}$~C\\
 Electronics parameters & \\
 \hspace{0.5 cm}Pulse shaping time & $1~\mu s$\\
 \hspace{0.5 cm}Dead time     & $5~\mu s$\\
 \hspace{0.5 cm}Number of channels in the spectrum  & 1024\\
 Mass & 1.35 kg \\
 Power & 6 W \\
 \hline
 \end{tabular}
 \end{table}

 At the heart of the instrument is a Silicon Drift Detector (SDD). Its unique configuration of electrodes provides very low detector capacitance and, thereby, higher spectral resolution than other silicon-based detectors that work in a similar energy range. 
 The SDD also has the ability to handle relatively higher incident flux. The detector, procured from KETEK GmbH, Germany, is available in the form of an encapsulated module containing a thermo-electric cooler (TEC), a FET, a temperature diode, and an $8~{\mu}m$ thick Beryllium entrance window.
 The detector has an active area of $\sim 30~{mm}^2$ and a thickness of $450~{\mu}m$.

 X-ray photons incident on the SDD generate a charge cloud proportional to 
 the deposited photon energy, which is collected at the anode of the detector. The front-end electronics 
 that include a charge sensitive preamplifier and shaping amplifiers 
 convert this charge to a voltage signal in the form of a semi-Gaussian pulse. 
 The peak of this pulse is then detected by a peak detector and digitized 
 by a 12-bit analog to digital converter (ADC). Histograms of the 10-bit ADC 
 channels (ignoring the two least significant bits), where each channel is 
 $\sim 16.5 ~eV$ wide, 
 are generated at an interval of one second and is recorded on-board.
  Apart from the complete spectrum with one-second cadence, the XSM also 
 records light curves in three pre-defined, but adjustable (by ground command), energy 
 intervals with a 100 ms time resolution. 
 XSM processing electronics packetizes the spectral data every second along with the
 health parameters of the instrument, such as the detector temperature, current drawn
 by the TEC, and various voltage levels. These packets are sent to the spacecraft
 data handling system that records it after tagging it with the onboard clock time.
 Absolute time is assigned to each data packet during ground processing by the correlation between
 onboard clock time and Coordinated Universal Time (UTC) from the available
 real-time telemetry.

 The spectral performance of the detector, defined by the energy resolution, 
 depends on its temperature.  In order to achieve the targeted resolution of better than 
 180 eV, the SDD temperature needs to be maintained at $\sim -35^{o}C$. 
 Since the ambient temperature during the lunar orbit is expected to vary 
 over a wide range between $-30^{o}C$ to $+20 ^{o}C$, the XSM employs a closed-loop temperature 
 control using the TEC that is part of the detector module, to maintain the 
 detector at $-35^{o}C$.
 The hot end of the TEC is interfaced with a thermal radiator to radiate away the heat.
 There is also a provision to vary the set point of the detector 
 temperature by ground command, in case such a requirement arises. 
    
 It is well known that the solar X-ray intensities vary over several orders of 
 magnitude between quiet and active phases of a solar cycle~\cite{2015A&A...582A...4J}. In the classification 
 of solar flares based on the flux measured by the GOES 1--8$~\AA$ channel, the highest 
 intensity X-class flares have five orders of magnitude higher flux than the lowest 
 intensity A-class flares. A single X-ray spectroscopic detector cannot be sensitive enough to detect 
 flares below A-class and, at the same time, does not saturate during large 
 flares. Hence, the XSM includes a filter wheel mechanism mounted on the top cover of the sensor 
 package that brings a 250$~{\mu}m$ Beryllium filter in front of the detector during 
 high-intensity flares. This additional Be filter increases the low energy cutoff and thereby reduces the incident 
 rate. An onboard algorithm moves the filter wheel to 
 the Be-filter position when the count rate is higher than the specified threshold for five 
 consecutive intervals of 100 ms. A similar automated logic decides the movement of the filter wheel mechanism 
 back to its open position. The threshold rate for movement to Be-filter position is nominally 
 set to be 80,000 $counts~s^{-1}$ (adjustable by ground command), which corresponds to $\sim$M5 class flare as discussed later.
 
 The filter wheel also has an additional position where a calibration source is mounted.                 
 The calibration source used is 100 mCi activity Fe-55 nuclide covered with a $3~\mu m$ thick titanium foil. 
 This source generates four mono-energetic lines: Mn-K$\alpha$ and Mn-K$\beta$ lines with energies of 
 5.9 keV and 6.49 keV, respectively, as well as Ti-K$\alpha$ and Ti-K$\beta$ lines of energies 4.5 and 4.93 keV, 
 respectively.   
 Spectral response and gain of the 
 instrument can be monitored by acquiring the spectrum of this calibration source. The 
 filter wheel mechanism can be brought to the calibration position by ground command for in-flight 
 calibration as and when required. 

 As the spacecraft attitude configuration is dictated by the requirement of observation of 
 the Moon by other instruments and other mission operation constraints, 
 the Sun's position varies with respect to the bore-sight of the XSM.
 Hence, in order to maximize the duration of observation of the Sun, the XSM is designed with a 
 large field of view of $40^{\circ}$ half cone angle. 
 An aluminium collimator (or detector cap) with a thickness of $0.5~mm$ placed over the detector provides 
 this wide field of view and, at the same time, restricts the aperture area 
 such that the count rate remains within the instrument capability over a wide range 
 of incident solar X-ray intensities.
 The collimator aperture of $\sim 0.7~mm$, much smaller in comparison to the detector diameter of $\sim 6.18~mm$, defines the instrument's geometric entrance area. 
 As aluminium with thickness $0.5~mm$ is transparent to X-rays above $\sim 8~keV$, the collimator is 
 coated with $50 ~{\mu}m$ of silver on both sides to block X-rays up to 15 keV arriving from outside the XSM FOV. 
 The coating on the bottom of the cap also ensures that the aluminium fluorescence lines from the cap 
 do not reach the detector.

 \begin{figure}
 \centerline{\includegraphics[width=1.00\textwidth]{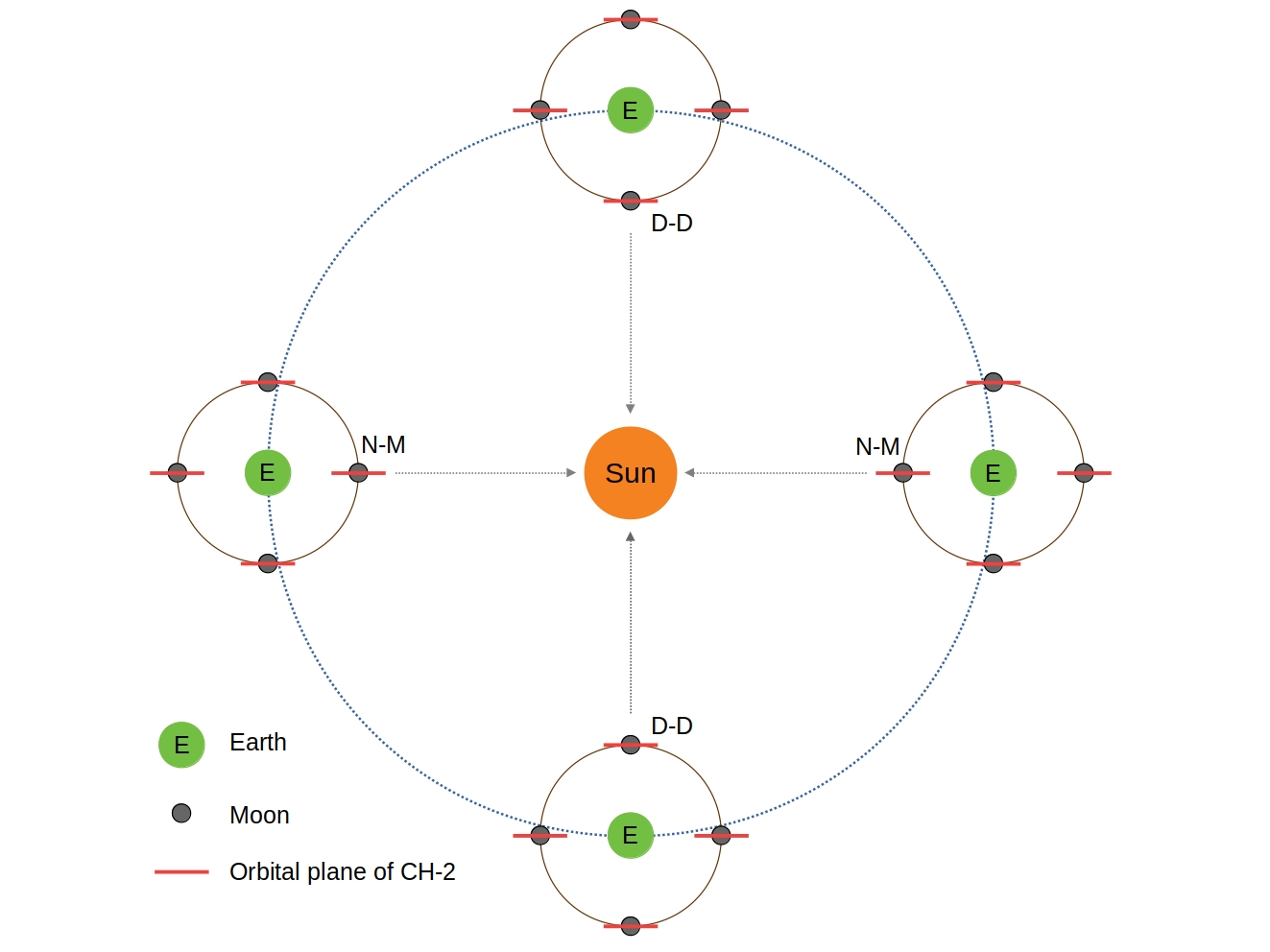}}
 \caption{ Orbital seasons for Chandrayaan-2 spacecraft: `dawn-dusk' (D-D) 
 and `noon-midnight' (N-M). On `dawn-dusk' day, the orbital plane 
 is perpendicular to the Moon-Sun vector, and on `noon-midnight' day, it is parallel. The attitude 
 of the spacecraft is defined differently during the three month seasons around both these 
 days, as described in the text.}
\label{orbital_season}
 \end{figure}

 During the nominal operation phase, the attitude of the  Chandrayaan-2 spacecraft is such that the yaw 
 direction always points to the Moon (see figure~\ref{xsm_mounting} for the axis definition). 
 All Moon-viewing instruments are mounted with their 
 bore-sights in the yaw direction. Thus the spacecraft attitude is not fixed in the inertial reference frame. 
 However, the orbital plane is fixed with respect to the inertial reference frame, and this results 
 in different orientation of the orbital plane with respect to the Sun giving rise to the 
 orbital seasons: `dawn-dusk' (D-D) and `noon-midnight' (N-M)~\cite{vanitha20}, as shown in figure~\ref{orbital_season}.
 The attitude of the spacecraft (other than yaw axis) is maintained according to these orbital 
 seasons to optimize power generation. 
 During the three months of D-D 
 season surrounding the day when the orbital plane is perpendicular to the Moon-Sun vector, 
 the spacecraft attitude is such that Sun is maintained in a yaw-roll plane. In this season, 
 continuous observations of the Sun with XSM are possible. In the  
 N-M season covering three months around the day when the Sun vector
 lies in the orbital plane of the spacecraft, the spacecraft attitude follows the 
 orbital reference frame, and the Sun does not remain within the FOV of XSM 
 for the entire orbit. Nominally, the XSM is expected to operate at all times when the Sun 
 is within the FOV of the instrument.

\section{Calibration requirements for XSM}
\label{xsm_gcalreq}

In case of an X-ray spectrometer like the XSM, the calibration primarily involves
the accurate determination of the response matrix which dictates the relation between
the incident photon spectrum, which is of interest, and the observed spectrum as:

\begin{equation}
C(I) = \int{N(E) ~R(E,I) ~dE}
\label{folding}
\end{equation}

\noindent  where $R(E,I)$
is the response matrix of the instrument, 
$N(E)$ is the incident spectrum of the source in units of $photons ~{s}^{-1}{cm}^{-2}{keV}^{-1}$, 
and $C(I)$ is the observed spectrum having units $counts~{s}^{-1}~{channel}^{-1}$.

The XSM records the number of events detected (counts) every second in
each ADC channel, also known as Pulse Height Analysis (PHA) channel. Gain parameters 
determine the mapping of the PHA channel to the nominal energy of the incident photon. 
As the gain usually varies with observing conditions, this mapping does not remain constant. 
In such a case, it is usual to resample the raw PHA spectrum into a pulse invariant (PI) channel 
space, correcting for the gain. This PI spectrum is used along with the response matrix 
to estimate the incident spectrum based on the relation given in equation~\ref{folding}.
We follow this approach for spectral analysis with the XSM.

Usually, the equation~\ref{folding} is approximated as a finite sum with $R(E, I)$ defined over
a grid of incident energies. This response matrix $R(E,I)$ can be divided into two 
components:

\begin{equation}
R(E,I)=\mbox{\textit{ARF}}(E)~ \mbox{\textit{RMF}}(E,I)
\end{equation}

\noindent  where $\mbox{\textit{RMF}}(E,I)$ is the redistribution matrix that incorporates the 
spectral redistribution effects due to the detector characteristics and $\mbox{\textit{ARF}}(E)$ 
is the ancillary response function (ARF) that includes the effective area of the instrument.

Thus, in order to infer the incident photon spectrum from XSM observations, gain parameters for 
the conversion of PHA spectrum to PI spectrum, the redistribution matrix, and the 
ancillary response function are to be determined during ground calibration. 
Considering these requirements, we investigate the linearity of the XSM over its entire energy range and 
then characterize the gain parameters over different operating conditions of temperature 
and incidence angle. The spectral redistribution function for the SDD used in the XSM is modeled using 
observed mono-energetic line spectra, and the effective area as a function of energy and angle 
is computed from experimental data and appropriate detector module parameters. 
We also study the stability of the spectral performance with flux, dead time, and pileup effects, which 
are of importance at high incident flux levels. 
In the subsequent sections, we describe these experiments carried out for the 
ground calibration of the XSM instrument and present the results.

\section{Gain calibration}
\label{gaincalsec}
 
The relation between the peak channel recorded by a spectrometer for mono-energetic 
lines and the incident photon energies provides the gain parameters. 
If this relation is linear then only two parameters: gain and offset 
are sufficient to describe it. 
These parameters are typically estimated by acquiring the spectra of sources 
with mono-energetic lines at known energies and obtaining the respective 
peak channel positions by fitting the observed spectra. 

 \begin{figure}
 \centerline{\includegraphics[width=1.0\textwidth]{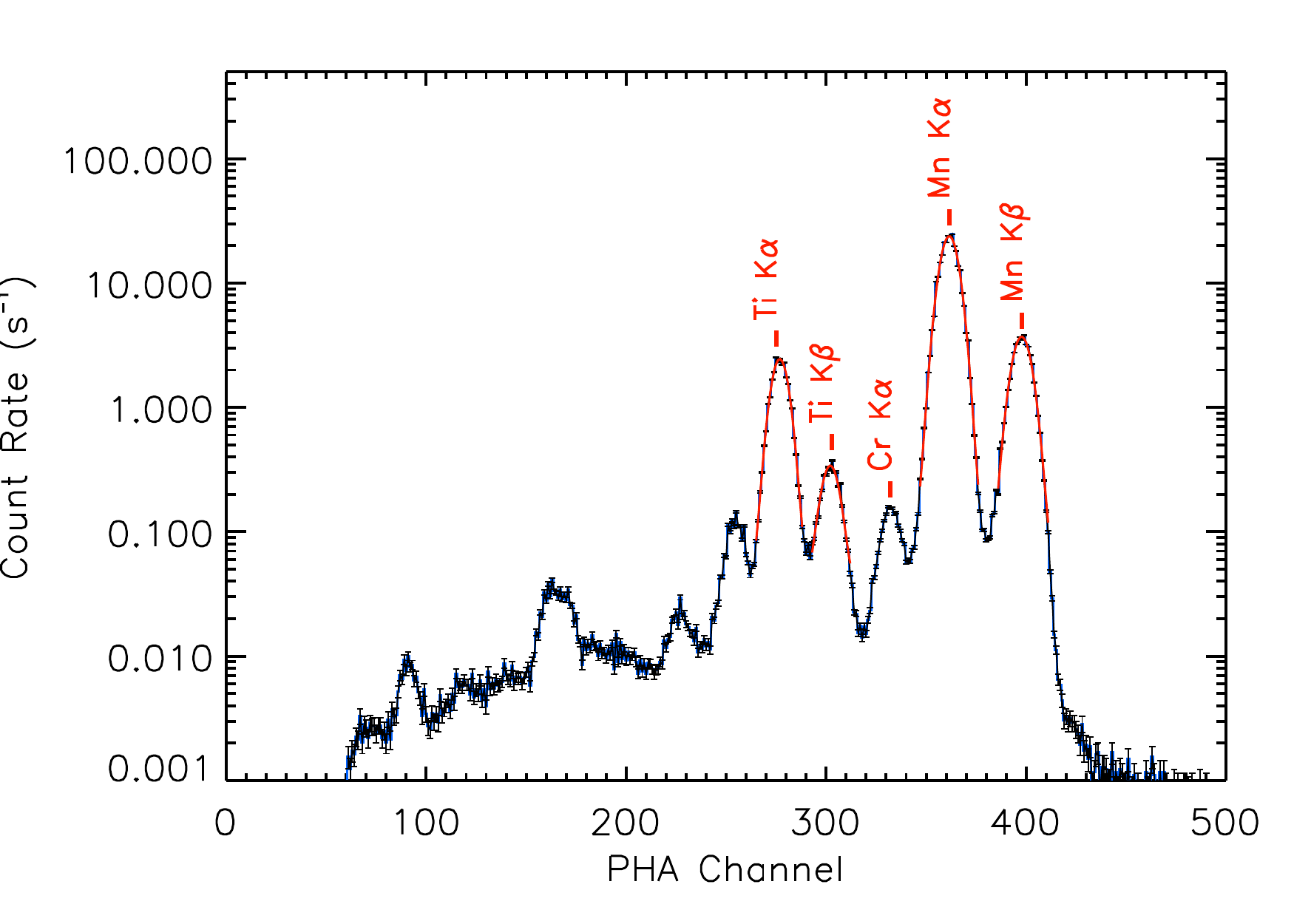}}
 \caption{A raw spectrum of the calibration source obtained using the XSM on the ground in PHA 
 channels (up to $\sim8~keV$) with the spectral lines identified. Four strongest lines were fitted with Gaussian to obtain 
the peak channels and width. The measured spectral resolution, defined by FWHM 
of the Mn$K\alpha$ line at 5.9 keV, is 175 eV.}
 \label{calsrc}
 \end{figure}

The XSM instrument includes a calibration source, which emits four major lines at known energies, 
mounted on the filter wheel and is used for gain calibration.
A spectrum of this calibration source obtained with the XSM at room temperature 
is shown in figure~\ref{calsrc}, where the major lines are identified. These lines were fitted with Gaussian models 
to obtain the peak channel position as well as the width. 
Peak channels and energies of these lines 
follow a linear relation.  The spectral resolution, 
defined by the full width at half maximum (FWHM) of the line at 5.9 keV, 
is measured to be $\sim$175 eV, which is better than the specified 
requirement of 180 eV. 

\begin{figure}
 \centerline{\includegraphics[width=1.0\textwidth]{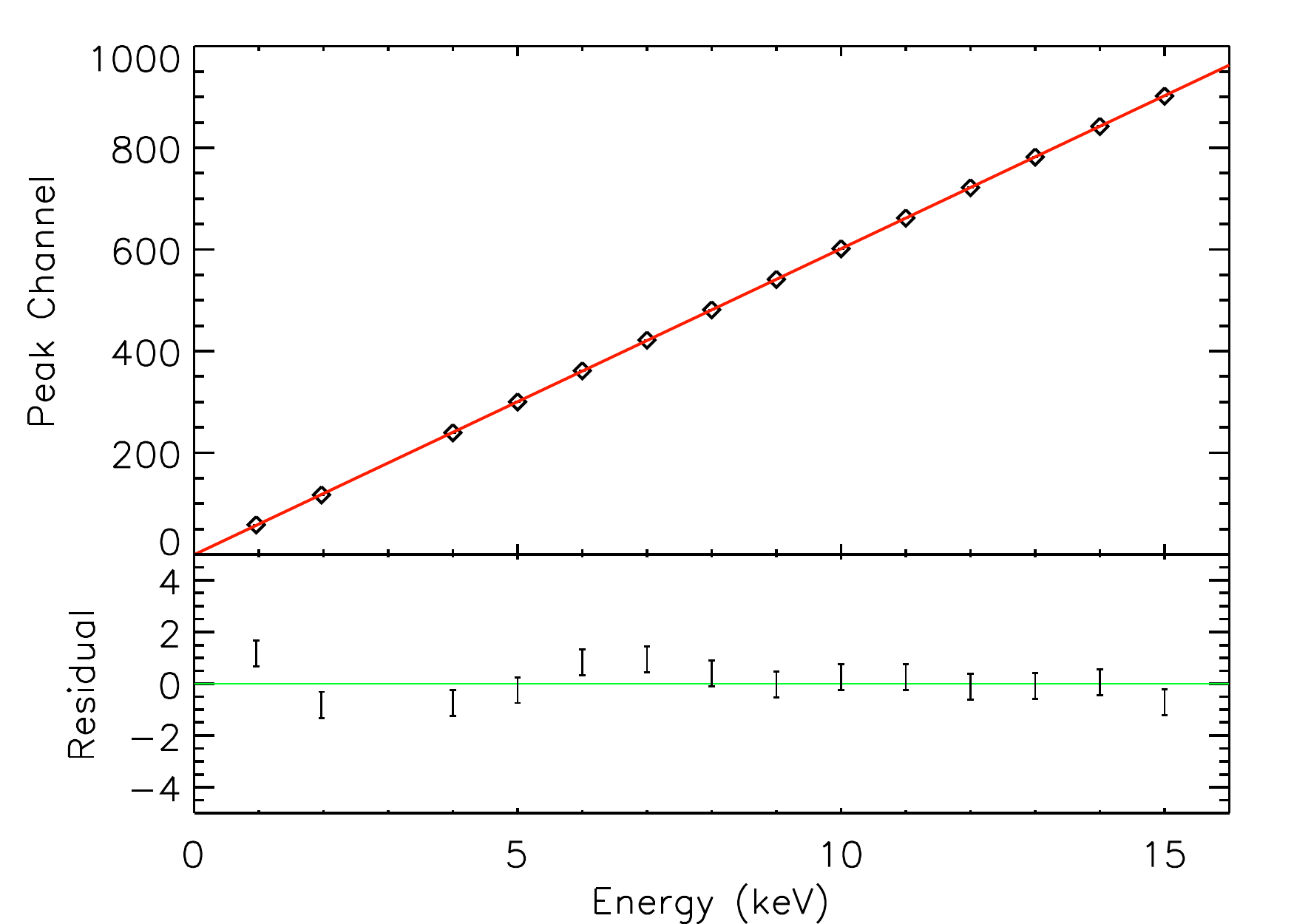}}
 \caption{Peak channel for mono-energetic lines plotted as a function of energy. Error 
bars are smaller than the symbol. The best fit line to the data is overplotted, and the residuals 
are shown in the bottom panel.}
 \label{linearity}
 \end{figure}

However, these measurements are restricted to only a small range of 4.5 -- 6.5 keV 
in the complete energy range of 1 -- 15 keV of XSM. Hence, in order to verify the 
linearity over the entire energy range and to model the response function (as discussed 
in the next section), experiments were carried out with XSM at BL-03~\cite{modi19} and BL-16~\cite{tiwari13} 
X-ray beam lines of 
Indus-2 synchrotron facility at the Raja Ramana Center for Advanced Technology (RRCAT), Indore. 
Spectra were acquired with the XSM of the mono-energetic X-ray lines in the energy ranges 
of 1 -- 2 keV (BL-03) and 4 -- 15 keV (BL-16). All measurements were carried 
out at low incident rates ($\sim< 5000~$ $counts~s^{-1}$) and at approximately 
the same temperature so that the spectral response is not affected by these 
two factors.  Observed X-ray lines were fitted 
with a Gaussian model to obtain the peak channel position
at respective incident energy, as shown in figure~\ref{linearity}, 
from which gain and offset are obtained with a straight line fit.
A systematic error of 0.5 channel ($~8 eV$) was added to the peak channel 
measurements to obtain a statistically acceptable fit.
Figure~\ref{linearity} shows that the XSM has a linear gain relation over its entire energy range 
and that it provides energy measurements correct within 10 eV.
Thus a linear fit with four lines from the in-built calibration 
source can provide the measurement of gain and offset of the instrument 
on the ground as well as during in-flight operations.

 \begin{figure}
 \centerline{\includegraphics[width=1.0\textwidth]{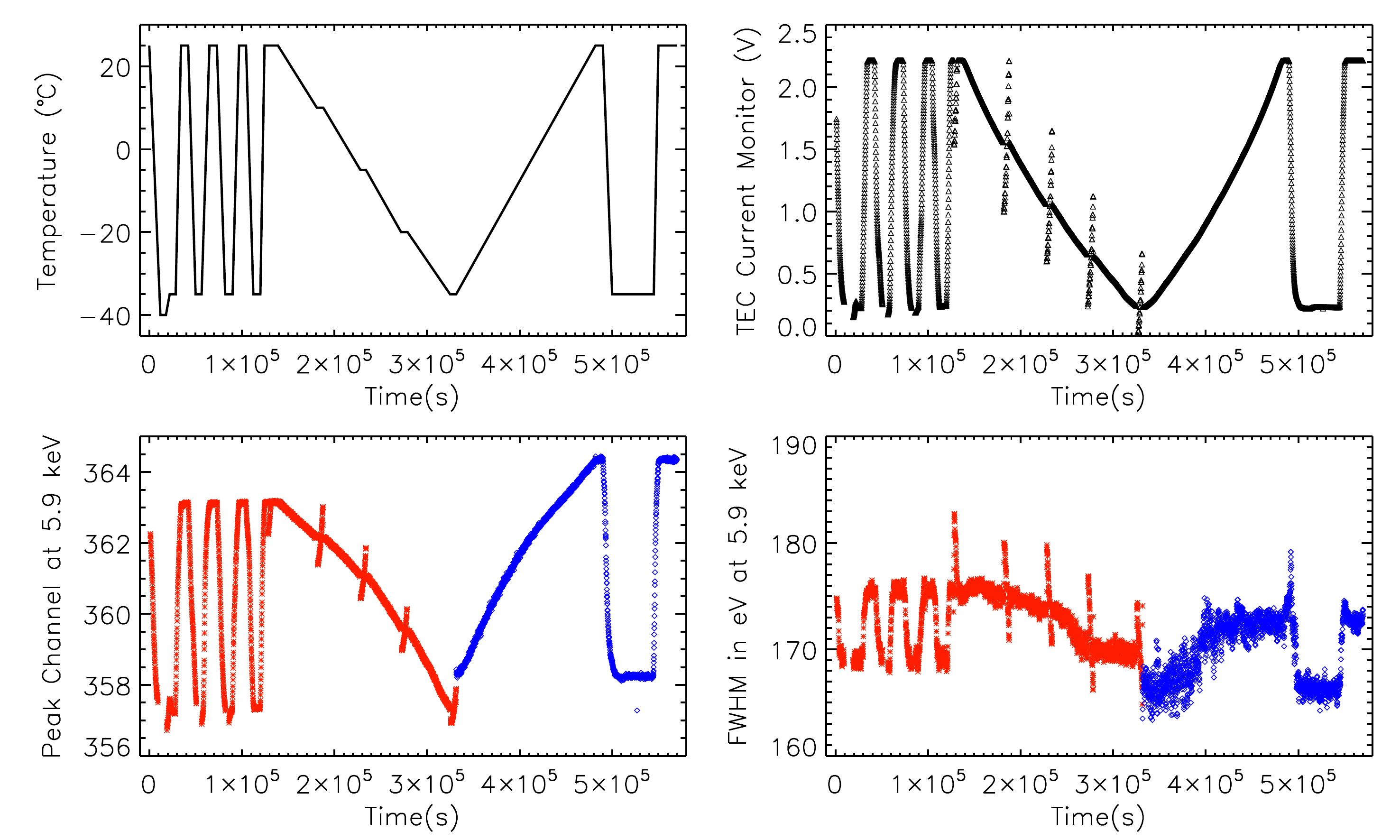}}
 \caption{Results of thermovaccum test of XSM are shown here. The upper two panels 
 show the XSM package temperature profile during the test (left) and the TEC current 
 monitor (right) corresponding to the same duration. The lower two panels show the 
 peak channel (left) and FWHM (right) of the spectral line at 5.9 keV.  
 Data obtained with the calibration source that is part of the instrument and 
 an external source are shown with the red star and blue diamond symbols, respectively. 
 During the slow downward ramp, the detector temperature was varied, keeping the package temperature 
 constant at five values.}
 \label{tvac_res}
 \end{figure}

Gain parameters of X-ray spectrometers often show variation with temperature.
Although the detector of XSM is maintained at a constant temperature, the instrument 
package temperature is expected to vary between -30$^{\circ}$~C and 20$^{\circ}$~C 
during the in-orbit operations.
As the components of the front-end electronics face this temperature variation, the 
gain may vary accordingly. Hence, the gain parameters have to be estimated 
as a function of the instrument temperature. For this purpose, the XSM spectra of 
radioactive sources  were acquired over the entire operating temperature (with 
additional margin) during the thermovacuum test of the instrument, maintaining the detector at its nominal operating temperature of -35$^{\circ}$~C. 
In the initial part of the experiment, the calibration source that is
part of the instrument was used, and for the latter part, an external source
of Fe-55 was used. 

Figure~\ref{tvac_res} 
shows the variation of the peak channel and FWHM of the 5.9 keV line for the 
entire duration. For comparison, the temperature profile and current drawn 
by the TEC of the detector that is proportional to the package temperature are 
also shown. Note that during the slow downward ramp of temperature, 
data were also acquired, keeping the instrument at a constant temperature but 
with the detector maintained at different temperatures in the range of -43$^{\circ}$~C 
to -23$^{\circ}$~C by changing the TEC set point.  
Measurements from the onboard source and the external source are marked with 
red and blue symbols, respectively, in the bottom panels of figure~\ref{tvac_res}.
It can be seen that there is a systematic variation in 
the peak channel (hence in the gain) with temperature. 
It can be noted that at the same temperature, peak channels for the 5.9 keV line 
from the onboard source and the external source are different; 
however, this is understood to be due to the dependence of gain on the 
interaction position of the photons in the detector and is discussed 
in detail in the subsequent paragraphs.     
    
\begin{figure}
 \centerline{\includegraphics[width=1.0\textwidth]{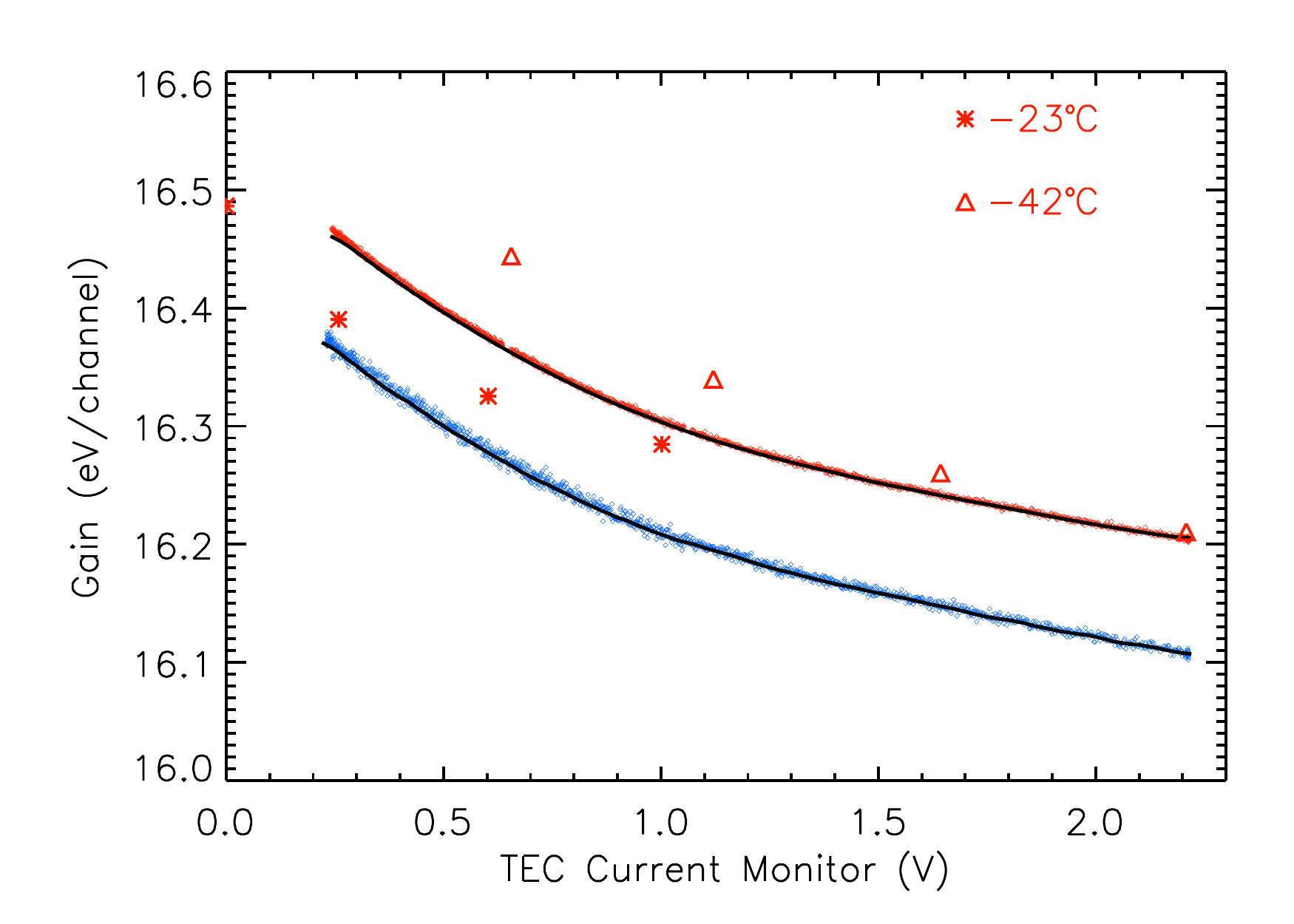}}
 \caption{The gain of the XSM obtained as a function of TEC current monitor for an external source 
 shining at the center of the detector (blue) and the onboard calibration source that 
 illuminates all regions of the detector (red) at the nominal detector temperature 
 of -35 $^{\circ}$~C. The black solid lines show the mean gain value over a grid 
 of TEC current values. The measured gain at two other detector temperatures of 
 -23$^{\circ}$~C and -42$^{\circ}$~C at a few instances of ambient temperature 
 are also shown.}
 \label{gain_offset}
 \end{figure}

For both cases, the peak channel - energy relations at different temperatures were fitted with a straight line. As the offset did not show 
any significant variation, it was frozen to the value obtained from X-ray beam line measurements, and the gain was 
obtained as a function of TEC current monitor that acts as a proxy for the instrument package temperature. 
Figure~\ref{gain_offset} shows the gain plotted against the TEC current 
for the case of the onboard source (red) as well as the external source (blue). 
In both cases, gain shows a similar monotonic dependence on the TEC current.
Measured values of gain at two other detector temperatures of 
-23$^{\circ}$~C and -42$^{\circ}$~C at few instances of ambient 
temperatures are also shown in the figure~\ref{gain_offset}. 
In case of any requirement to operate the detector at temperatures 
different from the nominal plan, gain parameters can be derived 
from similar measurements available.

As indicated earlier, another factor affecting the gain in the XSM is the 
interaction position of X-rays in the detector. The SDD used in the XSM has a
diameter of $\sim 6 ~mm$, whereas the aperture is of $\sim0.7 ~mm$ diameter.
When the source is on-axis to the instrument, the photons are incident
at the center of the detector. As the angle of incidence increases,
the interaction position on the detector moves away from the center
and gets closer to its edge.  With the increase in distance of interaction position from the 
center where the anode is located, the charge collection 
efficiency and hence the gain is expected to vary. The onboard calibration source is a disk of 
$\sim 6 mm$ diameter, and the placement is such that when it is in front of 
the detector, photons from the source will reach all positions of the 
detector. This explains the difference between the gain obtained using 
the calibration source and that obtained using the external source where the photons 
illuminate the center part of the detector alone.

 \begin{figure}
 \centerline{\includegraphics[width=1.0\textwidth]{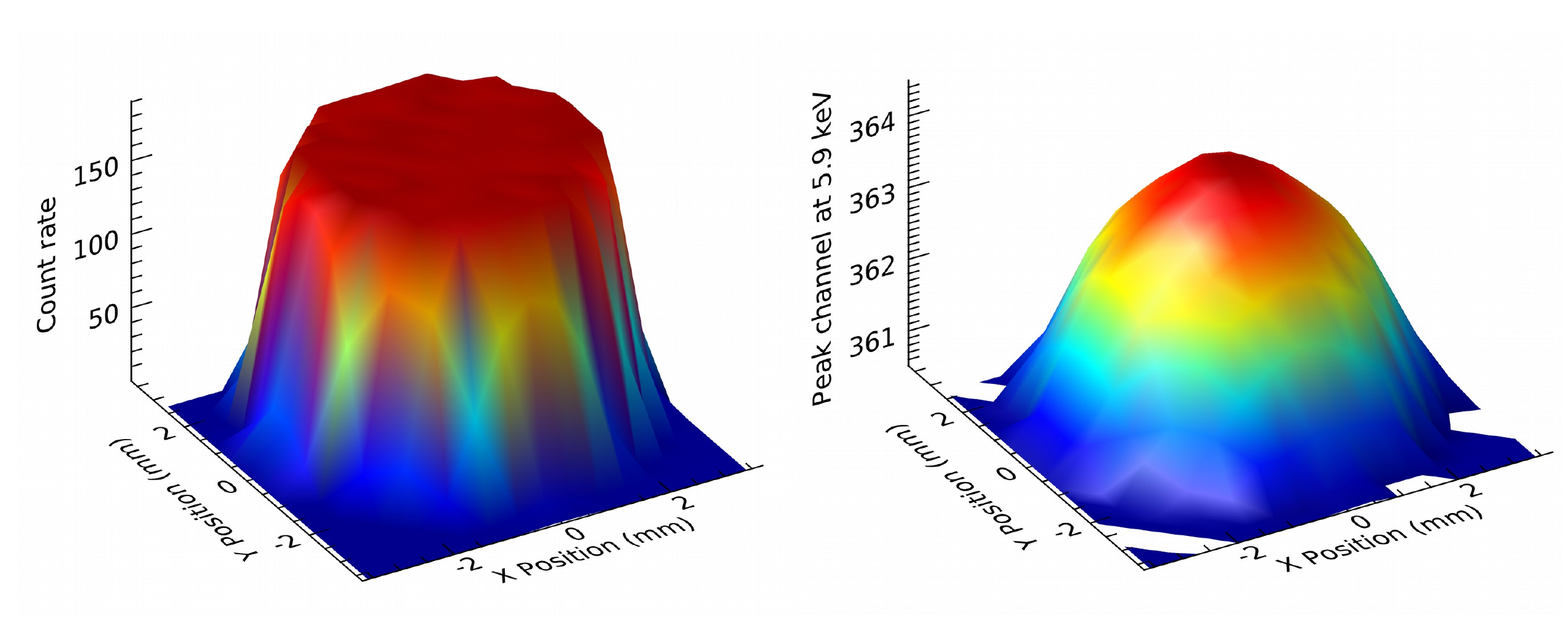}}
 \caption{Left: Surface plot showing the count rate over a grid of positions above the detector. 
 Right: Peak channel for the 5.9 keV line as a function of interaction position on the detector, 
 which shows a systematic trend. Colors are indicative of z-axis values.}
 \label{rate_pch_pos}
 \end{figure}

\begin{figure}
 \centerline{\includegraphics[width=1.0\textwidth]{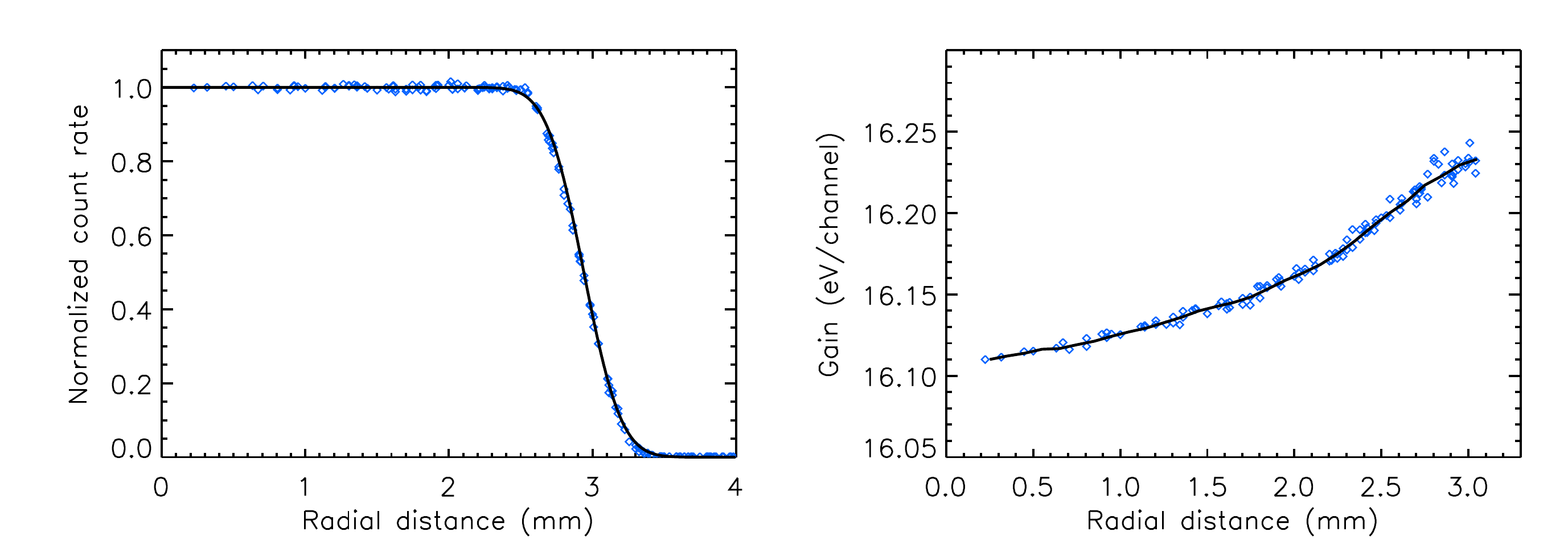}}
 \caption{ Left: Normalized count rate at each grid point as a function of radial distance. The black solid line 
 shows the expected trend when a Gaussian beam of $0.65 ~mm$ diameter 
 moves from entirely within the active area of the detector of radius  $3.09 ~mm$ 
 to entirely outside its active area.
 Right: Gain for each grid point shown as a function of the radial distance from 
 the center of the detector. The black solid line is the trend line obtained from the 
 data points.}
 \label{gain_pos}
 \end{figure}

In order to systematically investigate the dependence of the interaction position 
on the gain, 
an experiment was carried out where the detector without the 
collimator (detector cap) was illuminated at different positions with a pencil 
beam of diameter $\sim 0.7 ~mm$ from a radioactive source of Fe-55.  
The experiment was carried out at a constant ambient temperature. 
The beam was systematically moved over a regular grid of positions on the detector 
with a step size of $0.5~mm$ using motorized stages, and the spectrum was acquired in each case. 
Observed spectra were fitted with Gaussians for the lines at 5.9 and 6.5 keV 
to obtain their peak channel positions. 
Figure~\ref{rate_pch_pos} shows the surface plots of the total count rate 
and peak channel for 5.9 keV line over the entire two-d grid of positions.
It is seen that the  peak channel shows a small but definitive trend 
with the interaction position. 
Further, we compute the radial distance of each grid position from 
the center of the detector, and the count rate and gain at each position
are plotted against the radial distance in figure~\ref{gain_pos}.  The observed 
trend of count rates is consistent with the expected trend (overplotted with line) 
when a Gaussian beam of $0.65 ~mm$ diameter scans 
over the active area of the detector having a radius of $3.09 ~mm$, 
which confirms the manufacturer provided value of the detector active area.

 The results from the experiments discussed above 
show that XSM has a linear relation between energy and PHA channel, and 
the gain shows a systematic variation with the TEC current and 
the interaction position on the detector, which is determined by the Sun angle. 
As both TEC current and Sun angle will be available at every second in the in-flight data,  
accurate gain correction can be applied to the raw solar spectrum to convert 
it into the PI channels during ground processing. Further, any post-launch variation of the gain 
can be tracked using the onboard calibration source.

\section{Spectral redistribution function}

Photons of a given energy incident on an X-ray detector, are in general, recorded over a range of 
spectral channels of the instrument. This spectral redistribution occurs due to the inherent 
stochastic process of generation of the charge cloud, noise associated with the readout electronics, 
and other effects like incomplete charge collection. Spectral response of silicon detectors 
to mono-energetic X-rays typically consists of multiple components: a primary Gaussian photo-peak, 
an escape peak associated with Si fluorescence photons leaving the active detector volume, 
an exponential tail caused by incomplete charge collection, and a feature due to electron 
escape commonly referred to as shelf~\cite{scholze09}. 
The spectral redistribution function forms a major 
component of the response matrix of the detector system 
and hence required to be determined accurately for inferring the incident 
photon spectrum from the observed spectrum.  

\begin{figure}
 \centerline{\includegraphics[width=1.0\textwidth]{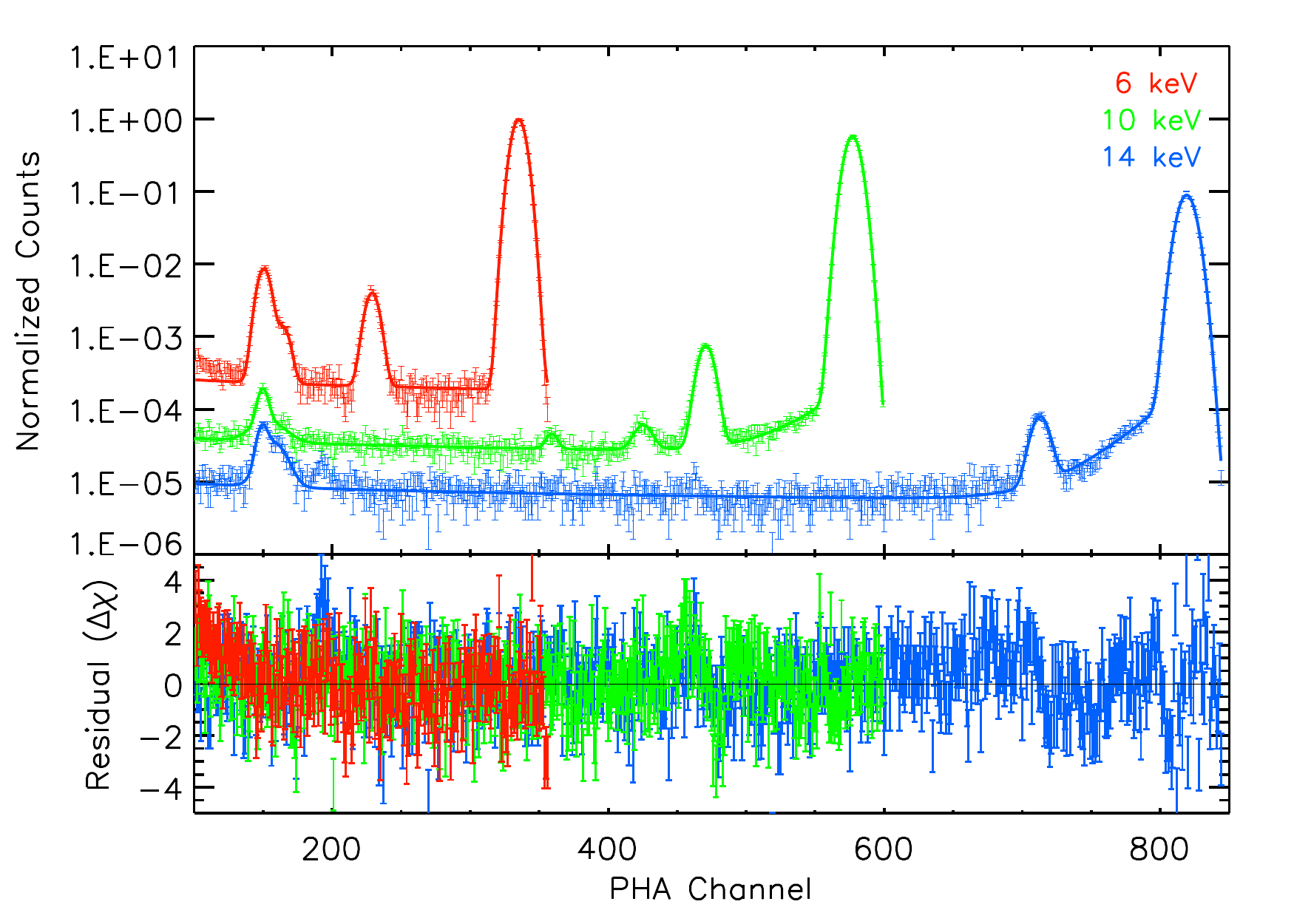}}
 \caption{XSM spectra of mono-energetic X-ray beams at three energies.  
 Note that for better clarity the spectra are shown with vertical offsets. 
 Best fit spectral response 
 models are overplotted with solid lines, and the bottom panel shows the residuals in terms of sigmas 
 with error bars of size one.}
 \label{rmf_fit}
 \end{figure}

In order to characterize the spectral response of the SDD, spectra of mono-energetic 
beams acquired with the XSM at RRCAT were utilized. 
Figure~\ref{rmf_fit} shows the spectra observed with XSM for three incident 
photon energies. 
The spectra of mono-energetic lines show all the expected features, 
as seen from the figure. 

We model the observed spectra for each energy with an empirical model that includes 
four components: 

\begin{equation}
C(E_i,E_0) = P1 + P2 + T + S
\end{equation}

\noindent where, $C(E_i,E_0)$ is the count spectrum in channel $I$ (with nominal energy $E_i$) for an incident 
photon energy of
$E_0$. The terms on the right hand side correspond to photo-peak ($P1$), escape peak ($P2$), 
exponential tail ($T$), and shelf ($S$). The primary and escape peaks are modelled with 
Gaussians defined as: 

\begin{equation}
P1 (E_i,E_0) = \frac{1}{\sqrt{2 \pi} \sigma} exp\Bigg[-\frac{{(E_i-E_0)}^2}{2 {\sigma}^2}\Bigg] 
\end{equation}

\begin{equation}
P2 (E_i,E_0) = I_{esc}~\frac{1}{\sqrt{2 \pi} \sigma} exp\Bigg[-\frac{{(E_i-(E_0-1.74))}^2}{2 {\sigma}^2}\Bigg]
\end{equation}

\noindent where, $\sigma$ is the standard deviation of the Gaussian and $I_{esc}$ is the 
relative strength of escape peak with respect to the primary peak. The tail component is 
modelled with an exponential function multiplied with complementary error function (erfc) as:

\begin{equation}
T(E_i,E_0) = I_{tail}~exp\Bigg[\frac{(E_i-E_0)}{\beta}\Bigg]*erfc\Bigg[\frac{(E_i-E_0)}{(\sqrt{2} \sigma)}+\frac{\sqrt{2} \sigma}{2.0 \beta}\Bigg]
\end{equation}

\noindent where, $I_{tail}$ is the relative strength of the tail component and $\beta$ 
is the parameter defining the slope of the tail. The shelf component in the model 
is given by:

\begin{equation}
S(E_i,E_0) = I_{shelf} {\Bigg(\frac{E_i}{10}\Bigg)}^{-\alpha} \Bigg(erf\Bigg[\frac{E_i-1.74}{\sqrt{2} \sigma}\Bigg]+2.0\Bigg) \mbox{, when } E_i < E_0 
\end{equation}

\noindent Here, the relative strength of the component is determined by $I_{shelf}$ and the 
parameter $\alpha$ is the power-law index.  

This empirical model for the SDD spectral response was implemented as a local model 
in pyXspec, the python interface of the spectral fitting tool XSPEC~\cite{arnaud96}. The model 
parameters include the relative strengths of all components, the width of the 
Gaussian ($\sigma$), tail slope $\beta$, and shelf power-law index $\alpha$.  
The observed spectra for all energies were fitted with the model. 
Additional line components were added in the model in cases where peaks, such as that of argon arising from the air in between, are present in the 
spectrum, to obtain proper fit (see figure~\ref{rmf_fit}). 
During the fitting, it was noted that the parameters $\alpha$, $\beta$, 
and $I_{shelf}$ are independent of energy, and they were frozen to 
constant values to obtain the final best fit values for 
rest of the parameters. Best fit models are overplotted on the 
observed spectra shown in figure~\ref{rmf_fit}. 
Residuals plotted in the bottom panel of the figure
show that the model describes the observed spectra well. 

 \begin{figure}
 \centerline{\includegraphics[width=0.7\textwidth]{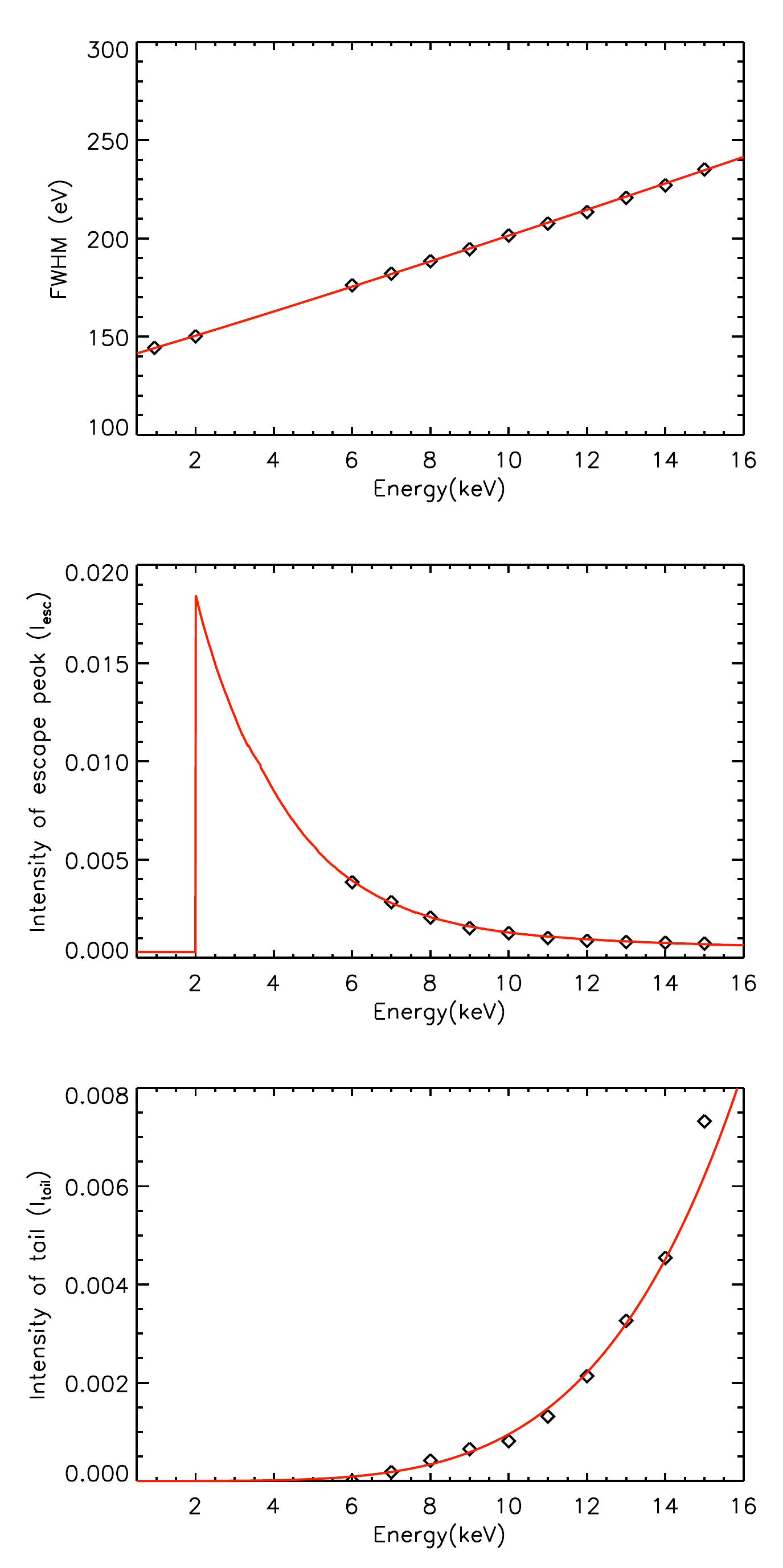}}
 \caption{Spectral response model parameters obtained from fitting mono-energetic spectra: 
 FWHM (top), the relative intensity of the escape peak (middle), and intensity of the tail 
 component (bottom) are plotted as a function of energy. Statistical errors are smaller 
 than the symbol size. The red solid lines show the best fit models given by 
 equations~\ref{fwhm_eqn} and ~\ref{tail_eqn}, in the case of FWHM and intensity of the 
 tail component, respectively, whereas for the escape peak intensity (middle panel), the line 
 represents the result from Geant4 simulation.}
 \label{rmf_par}
 \end{figure}

In order to derive the spectral response of the instrument over the 
full energy range of 1 -- 15 keV, the model parameters are required 
as a function of energy. Figure~\ref{rmf_par} shows the three free 
parameters of the model plotted as a function of incident photon energy. 
The full width at half maximum (FWHM) of the Gaussian, defined as $2.35 \sigma$, 
to a first-order has a dependence on the square root of energy. However, the best 
fit is obtained with the inclusion of a second-order term using 
a function of the form: 

\begin{equation} 
\mbox{FWHM} ~(E) = 2.35~\sqrt{a+b E + c E^2}
\label{fwhm_eqn}
\end{equation}

\noindent  In this equation, FWHM has units of eV and E is in keV. 
The best fit curve is overplotted with the data points in the figure~\ref{rmf_par}. 

 The second panel of the figure~\ref{rmf_par} shows the relative intensity of the escape peak as obtained 
from the spectral fits. As escape peaks are present only when incident photon energy 
is higher than the K-edge energy of Si, this parameter is not applicable for 
the monochromatic line at 1 keV. For the line at 2 keV, the escape peak energy is 
well below the lower energy threshold of XSM. Hence the measurement of escape 
peak strength is available only for 6 keV and above. As the measurements are not available near to the K-edge 
energy, we resort to simulations with Geant4 toolkit~\cite{agostinelli03} 
to obtain escape peak strength over the entire energy range. 
A silicon 
detector is modeled in Geant4 with the dimensions matching the detector used 
in XSM. The detector is illuminated with a large number of photons of each photon energy, 
and energy deposited is recorded. From the simulation output, the intensity of the 
escape peak is computed, and the result is overplotted as a solid line in figure~\ref{rmf_par}, 
which is found to match with the measured values at few energies.       
 The third free parameter, the intensity of the tail component, shows an increasing 
trend with energy. It is fitted with a power-law model: 
\begin{equation}
I_{tail} (E) = I_{tail}^{0}~{\Bigg(\frac{E}{15}\Bigg)}^{{\gamma}}
\label{tail_eqn}
\end{equation}
 The best fit model overplotted on the data is shown in the bottom panel of the figure~\ref{rmf_par}. 

 \begin{table}
 \caption{ Parameters of spectral redistribution function model of XSM defined by equations 3-9.}
 \label{tab_rmfpar}
 \begin{tabular}{l l}
 \hline
 Parameter & Value \\
 \hline
 $a$  & 3528.9 \\
 $b$  & 299.04 \\ 
 $c$  & 9.41 \\
 $I_{esc}$  &  Geant4 Simulation\\
 $I_{tail}^{0}$ & $6.21\times10^{-3}$\\
 $\gamma$  & 4.63\\
 $\beta$   & 0.5\\
 $I_{shelf}$ & $10^{-4}$\\
 $\alpha$ & 0.3\\
 \hline
 \end{tabular}
 \end{table}

Thus, we obtain all the parameters of the spectral redistribution function 
model, as tabulated in table~\ref{tab_rmfpar}. For FWHM and intensity of 
the tail component, parameters defining the energy dependence in equations 
~\ref{fwhm_eqn} and ~\ref{tail_eqn} are given in the table.
With these parameters, we evaluate the spectral redistribution 
function at a discrete set of energies to generate
the redistribution matrix of 
XSM, which is required in spectral analysis. 
It is also noted that specific PHA channels (typically channel numbers ${2^n}$ with $n\ge4$) have 
systematically higher or lower counts than expected from the redistribution model, 
and this is attributed to the ADC characteristics. Systematic errors for each PHA 
channel were estimated from the ground calibration data, and typical errors are 
$\lessapprox 2\%$ with a maximum up to 10\% for a few channels. These channel-wise 
systematic errors are incorporated into the errors on the spectral measurements.
As all components of the mono-energetic response of the detector and electronics 
are incorporated in the model, it provides an accurate redistribution matrix.

\section{Ancillary response function}
\label{gcal_ARF}
 The ancillary response function (ARF) or the effective area of XSM includes 
geometric area, collimator response, and detection efficiency and 
transmission of entrance windows as a function of energy. 
As the angle subtended by the Sun with respect to 
the axis of XSM is expected to vary, the effective area needs to be estimated 
as a function of incidence angle. 

\begin{figure}
 \centerline{\includegraphics[width=1.0\textwidth]{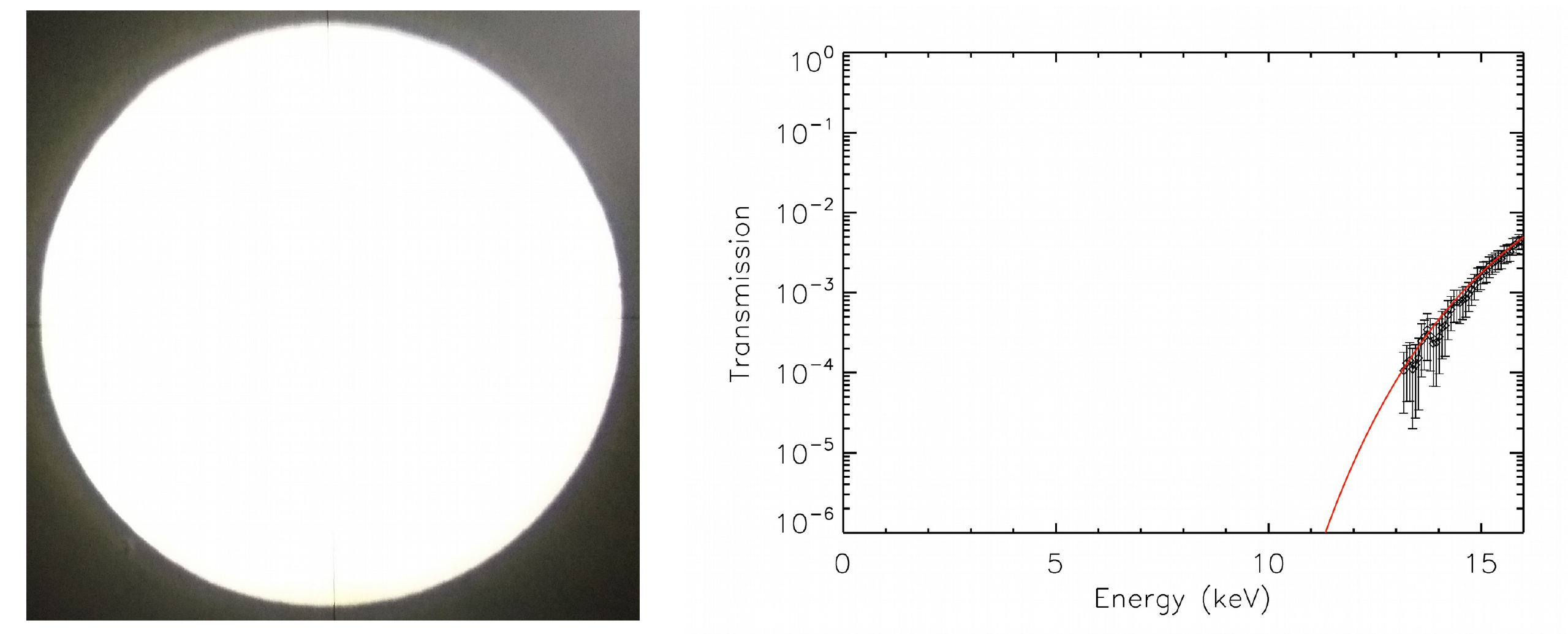}}
 \caption{ Left: Photograph showing the aperture of the XSM collimator. Right: Transmission 
 of the silver-coated ($50~\mu{m}$ on both sides) collimator obtained as the
 ratio of X-ray gun spectrum recorded  with specially made collimator without any aperture to that recorded without the 
 collimator. 
 The red solid line shows the expected transmission 
 of the collimator.}
 \label{detcap}
 \end{figure}

The primary component of the effective area 
is the geometric area, which in the case of the XSM, is defined by the aperture 
on the aluminium collimator (detector cap).  
The collimator is designed to have an aperture of 0.7 mm; however, it is essential to measure the fabricated unit's aperture as it may deviate from the design value.
As the aperture 
size may vary between the multiple fabricated units, measurements 
of diameter across various directions were 
carried out for all of them with an optical projection facility.   
The unit with the least non-uniformity in aperture was used for the 
flight model of the instrument (photograph shown in figure~\ref{detcap} left panel).
For this unit, the diameter of the aperture was measured to be $0.684 ~mm$, 
which defines the on-axis geometric area of the instrument. 

The efficacy of the aluminium collimator, which is coated with $50 ~\mu{m}$ silver on both sides, 
to block the X-rays from all directions other than the aperture was also verified experimentally. 
A specially made silver-coated collimator without any aperture was assembled in front of the detector, 
illuminated with X-rays from a miniature X-ray generator (Amptek Mini-X with a gold target) and 
the spectrum was recorded. 
Spectrum without the collimator directly exposing the detector was also recorded, and
from the ratio of both these spectra, the transmission of the collimator was obtained as a function of energy, 
which is shown in the right panel of figure~\ref{detcap}.
This demonstrates that the collimator efficiently blocks X-rays over the 
energy range of XSM as required.

 \begin{figure}
 \centerline{\includegraphics[width=1.0\textwidth]{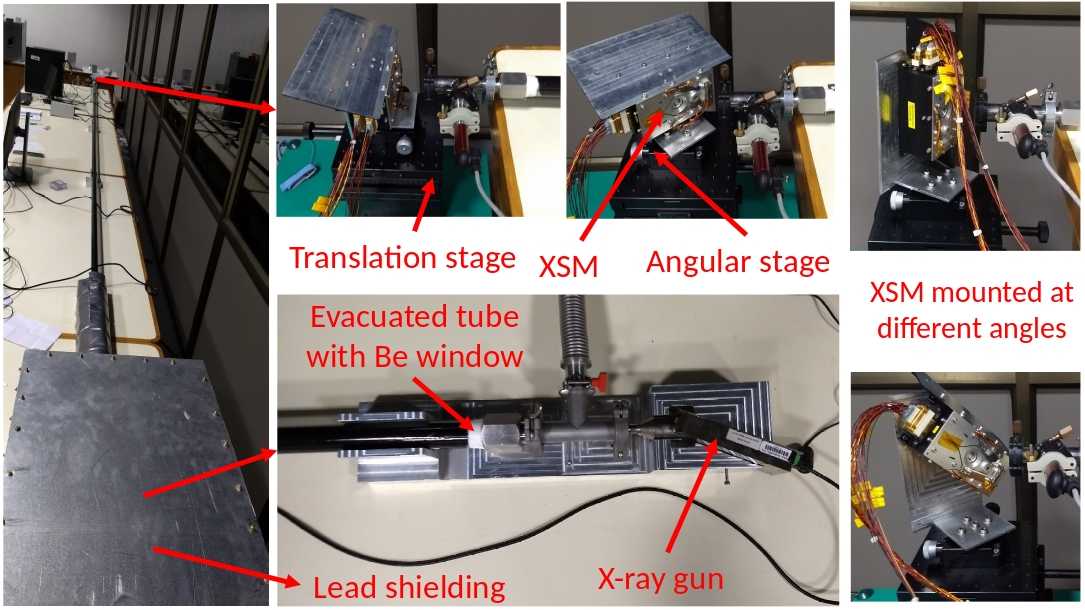}}
 \caption{Experimental setup for characterization of the angular response of the XSM. X-rays 
 from a miniature X-ray source was collimated over a distance of six meters using 
 an evacuated tube or beam line.  
 At the other end of the beam line, the XSM sensor package was mounted on rotation and translation stages. 
 XSM data acquired at different angles of the instrument with the incident beam 
 were used to obtain the relative area as a function of angle. Measurements were also 
 carried out with the sensor package mounted in different orientations, as shown 
 in the right-most images to determine the response across different azimuthal 
 directions.}
 \label{angresp_setup}
 \end{figure}

To determine the next component of ARF, the collimator response as a function of incidence angle, 
an experiment was carried out with a six-meter long temporary beam line, as shown in figure~\ref{angresp_setup}. 
 The beam line consisted of an evacuated steel tube with Beryllium windows on both ends to 
allow entry and exit of X-rays. 
A miniature X-ray source was placed on one end, and the XSM sensor package was mounted on translational and angular stages at the other end of the beam line.
The design of the mounting setup ensured that the rotation of the 
sensor package was with respect to the center of the aperture. Before the 
measurement of angular response, the X-ray beam was aligned to the XSM aperture 
by adjusting the position of the sensor package with the translation stage.
Data were acquired with the XSM at different angles of the instrument Z-axis (detector normal) with 
respect to the beam direction. A similar exercise was carried out by mounting 
the XSM in three more orientations to obtain the response across different azimuthal angles.

 \begin{figure}
 \centerline{\includegraphics[width=1.0\textwidth]{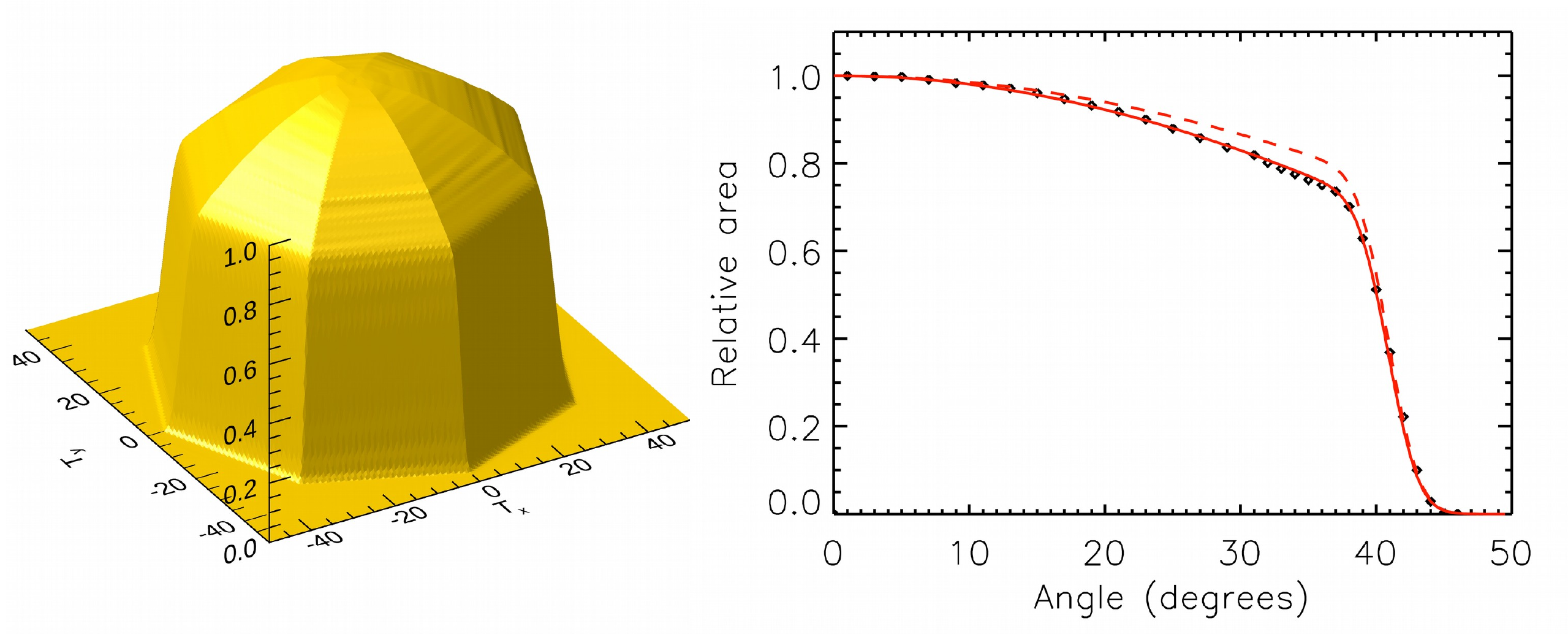}}
 \caption{Left: Surface plot showing relative area measured as a function of polar angle at eight 
 azimuthal angles. Right: Area as a function of polar angle at an azimuth angle of zero degrees. The red 
 dashed line shows the trend expected with the projection effect, and the red solid line is an empirical 
 model that explains the observed behavior.}
 \label{angresp_res}
 \end{figure}

Count rates at each angular position were determined from the observed spectra, and the 
relative area as a function of polar and azimuthal angles (with respect to the instrument 
reference frame defined in figure~\ref{xsm_mounting}) was estimated.  
Figure~\ref{angresp_res} left panel shows the three-dimensional surface plot of the angular response of XSM 
across eight azimuthal angles. 
 These measurements show that the field of view of XSM is symmetric, and the null points 
of the FOV are approximately at 44 degrees.
The right panel of the figure shows the angular response 
with polar angle along the azimuthal angle of zero degrees. The red dashed line overplotted 
on the figure shows the expected variation in angular response, considering only the 
projection factor along with the edge effects, and the data points are clearly lower 
than the prediction. Further, we model the observed angular response with a function 
having a $cos^{\alpha}(\theta)$ dependence instead of $cos(\theta)$ dependence, and 
the model is shown with a solid red line in the figure. It is seen that the observed 
angular response is explained with an index of $\alpha \approx 1.4$. 
The departure from the expected cosine behavior is attributed to the 
fact that the aperture is not a two-dimensional circle, but a 
three-dimensional cylinder of finite thickness 
of the order of $50~\mu m$. 

Other factors that contribute to the effective area are the thicknesses of the detector chip, 
dead layers, the Be entrance window, and the Be filter on the filter wheel mechanism. 
An attempt was made to measure the parameters that are internal to the detector module
with beamline experiments; however, there were very large uncertainties due to 
various other factors. Hence, we decided to use the manufacturer provided values for the 
thicknesses of the detector chip ($450~\mu m$), deadlayers ($0.1~\mu m$ Si and $80 nm$ SiO$_2$), 
and Be entrance window ($8~\mu m$). 
For the additional Be filter, physical measurement of the thickness, as well as the measurement of the transmission 
of X-rays, were carried out to obtain the final value of $254~\mu m$, which is within 
limits specified by the manufacturer. The overall detector efficiency as a function 
of energy is then estimated for different off-axis angles considering the transmission 
by the deadlayers, entrance window, and filter and the absorption by the silicon chip using the 
attenuation coefficients of the materials from the NIST database.      

 \begin{figure}
 \centerline{\includegraphics[width=0.9\textwidth]{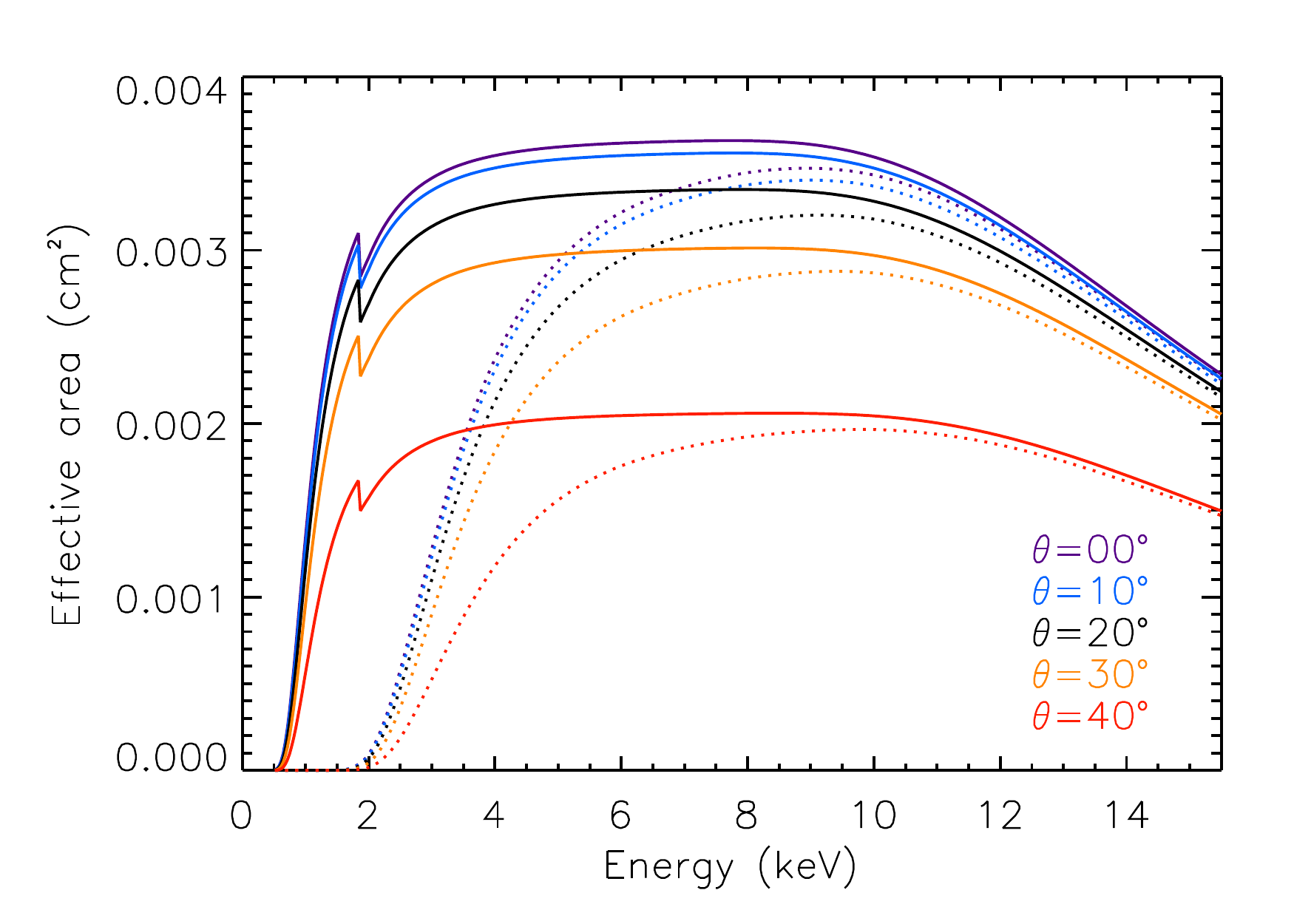}}
 \caption{The effective area of the XSM as a function of energy at different angles of incidence. Solid 
 lines correspond to the effective area without the Be filter, whereas the dotted lines show 
 the effective area with the Be filter in front of the detector.}
 \label{effarea_angene}
 \end{figure}

 Taking into account the geometric area, collimator response, and detector efficiency, 
the estimated effective area of the XSM at different incidence angles 
are shown in figure~\ref{effarea_angene}. Solid lines correspond to the effective area 
without the additional thick Be filter, whereas the dotted lines correspond to 
the cases with the filter. 

\section{Spectral performance with incident rate}
\label{crate_sec}
Since the incident count rate of the solar X-rays is expected
to vary over a wide range, one important aspect to examine 
for a solar X-ray spectrometer
is the stability of spectral performance with count rate.
X-ray spectrometers with analog front-end electronics typically
exhibit a change in the peak channel
position and degradation of spectral resolution with an increase in count rate.

In order to characterize the performance of the XSM with incident count rate, two sets
of experiments were carried out:  one with mono-energetic lines from X-ray beams at
RRCAT and another with a miniature X-ray generator.
XSM spectra were acquired for beam energies of 5 keV and
12 keV for a range of incident count rates. The count rates were varied with
the help of slits available in the beam line. 
 
With the miniature X-ray generator, XSM spectra were acquired for different 
count rates by placing the source at different distances from the instrument.
Distances were chosen such that the count rates span from few thousand $counts~s^{-1}$ 
to $\sim 4\times 10^5$ $counts~s^{-1}$.
X-ray spectrum from the gun includes
a Bremmsstrahlung continuum and L lines of the target material gold.
The spectral performance is evaluated with the $L-\alpha$ line of gold
at 9.7 keV.

\begin{figure}
 \includegraphics[width=0.9\textwidth]{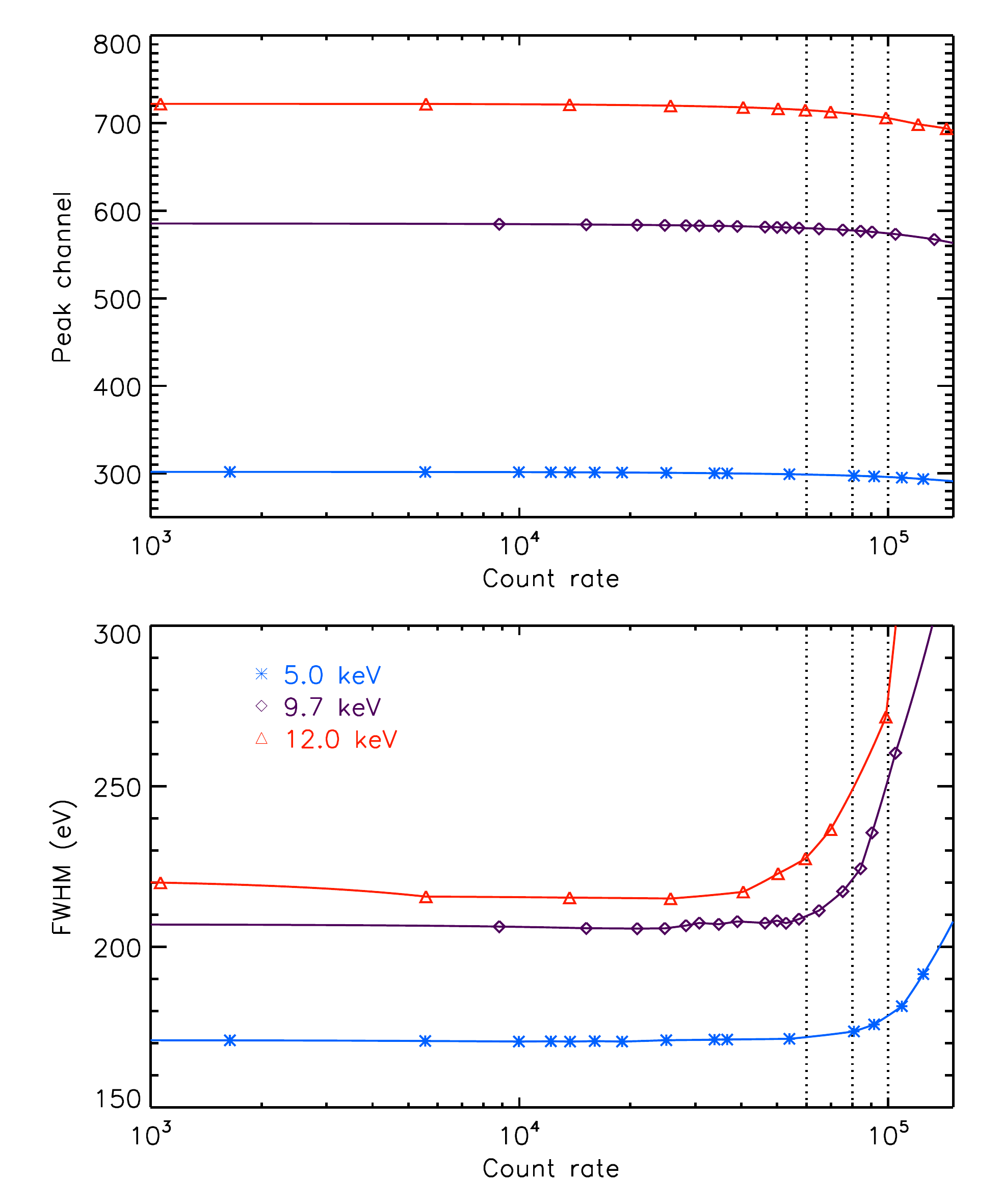}
 \caption{Peak channel (top) and FWHM (bottom) of spectral lines at three energies
 measured with the XSM as a function of incident rate. Vertical dotted lines
 show the approximate count rate above which the spectral performance degrades significantly
 for all cases. For the 9.7 keV line, spectral performance is stable up to an incident rate of 80,000 $counts~s^{-1}$.}
 \label{pch_fwhm_withrate}
 \end{figure}

For all three cases, spectral lines were fitted with Gaussian to obtain
the peak channel position and the FWHM. Figure~\ref{pch_fwhm_withrate}
shows these parameters for three energies plotted against the
event rate.
It can be seen from the figure that for all three energies, the peak channel
and FWHM show variation above threshold count rates demarcated by vertical
dotted lines. As one would expect, the threshold rate is the highest for 5 keV, 
the lowest of three energies, and decreases at higher energies. As most of the events in the solar
spectrum would be at lower energies, it is expected that spectral performance
during solar observations would be closer to the 5 keV case where
there is no significant degradation up to $10^5$ $counts~s^{-1}$.
However, to be on a safer side, we consider that the spectral
performance is stable up to $8\times10^4$ $counts~s^{-1}$ and have finalized
this as the threshold rate for automated movement of the filter wheel
to the Beryllium filter position.

Simulations using the XSM response matrix and the CHIANTI atomic database~\cite{1997A&AS..125..149D,2015A&A...582A..56D} 
to estimate the expected count rate for different levels of solar activity, 
details of which will be reported elsewhere, 
show that for M5 class of solar flares, the XSM will detect $\sim8\times10^4$ 
$counts~s^{-1}$. Hence, the transition from open to the Be filter position 
is expected at this level of activity. With the Be filter, the limiting 
rate of $8\times10^4$ $counts~s^{-1}$ will be reached during X5 class of flares, beyond 
which the spectral performance will begin to degrade.

 \begin{figure}
 \includegraphics[width=0.9\textwidth]{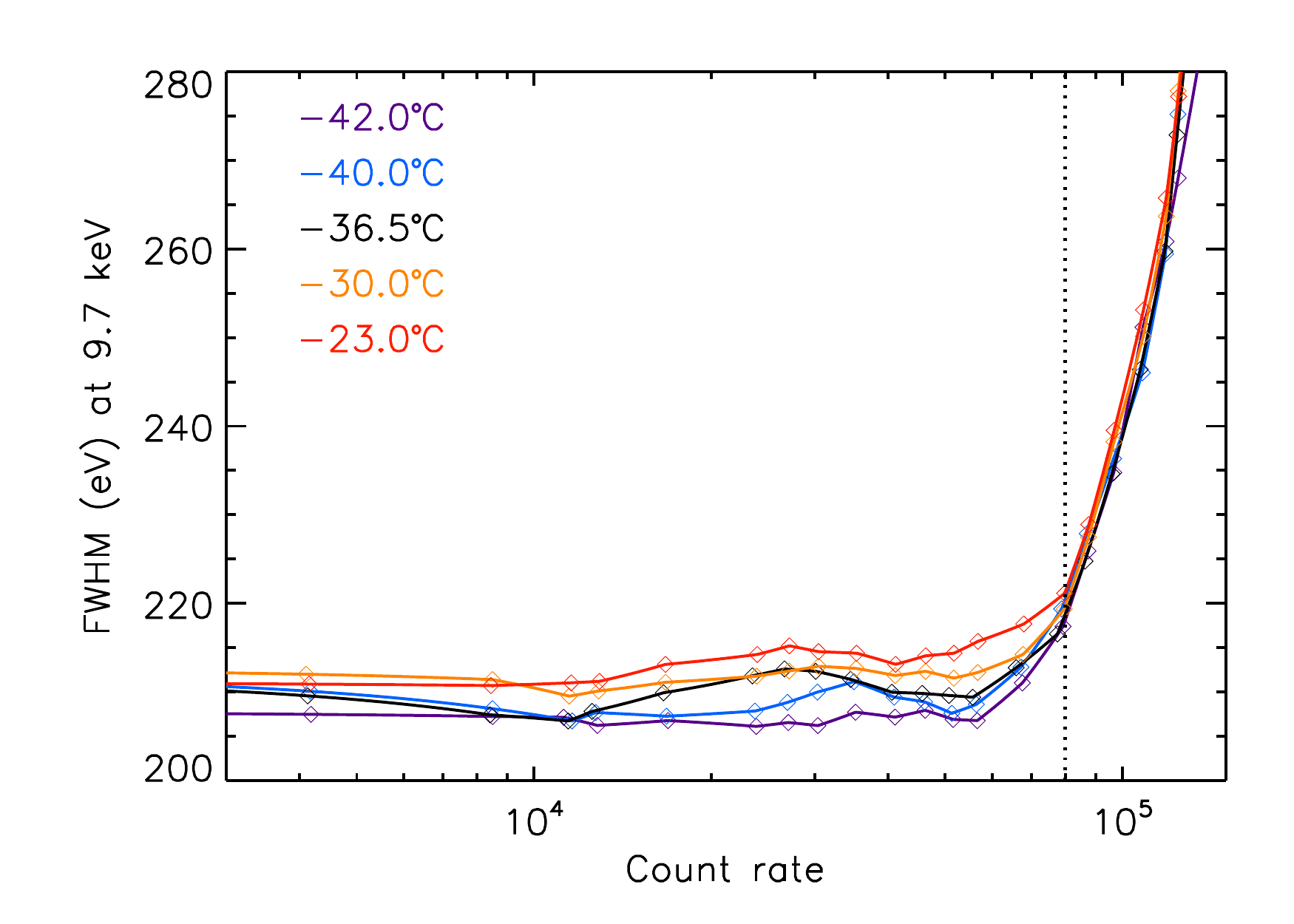}
 \caption{FWHM at 9.7 keV as a function of count rate with the detector maintained at
  different temperatures. The FWHM increases with the detector temperature; however,
  it can be seen that the count rate at which the performance starts to degrade is not significantly
  dependant on the detector temperature.}
 \label{fwhm_rate_dettemp}
 \end{figure}

We also examined the stability of spectral performance with the incident rate at different
detector temperatures. This was done by acquiring spectra from the miniature
X-ray source with different count rates and setting the detector temperature at specific
values. Figure~\ref{fwhm_rate_dettemp} shows the variation of FWHM with count rate
at different detector temperatures. The constant value of FWHM at lower count rates
shows a systematic increase with the temperature; however, the threshold event
rate where the FWHM degrades significantly does not seem to depend on the
temperature. Hence, even if there is a requirement to operate the XSM at higher detector
temperatures, it will not affect the dynamic range of the instrument.

\section{Dead time and pileup}
\label{deadtimesec}
 All photon counting systems have a short but finite dead time. Once an X-ray 
photon interacts,  it cannot sense another photon interaction while 
the detector and the front-end electronics process the signal from the first photon. 
The effects of dead time, which is typically of the order of few microseconds,
are important when the incident photon rate is high, as expected during 
high-intensity flares for XSM. To obtain accurate flux from the observed spectrum, 
the exposure times are to be corrected for the dead time effects. 

The XSM front-end electronics require $\sim 3-5 ~\mu s$ to properly digitize the height 
of the electronic pulse generated by any photon interaction. The first 
microsecond is an absolute dead time, during which two independent photon 
interactions cannot be distinguished separately. If the second photon 
interacts after one microsecond but within the digitization dead time, 
it can be identified as an independent interaction (counted as event triggers), 
but its energy measurement is not possible. The exact dead time of the analog 
front-end electronics depends on the energy of the incident photon and hence 
is difficult to correct. Therefore, XSM implements a fixed dead time of 
$5~\mu s$ (default value, can be changed by a command to $10~\mu s$ if needed) 
in the FPGA based digital readout system. Further, the XSM implements this fixed digital dead time in
a \textit{paralyzable} mode~\cite{knoll00}, i.e., event triggers that occur within the 
dead time are not considered for energy measurements, but they extend the 
dead time further. 
The event triggers themselves do not represent the actual incident 
rate as they are also affected by the absolute dead time of $\sim 1 ~\mu s$. 
However, in this case, the dead time behavior is expected to be similar to the 
\textit{non-paralyzable} model.

\begin{figure}
 \includegraphics[width=1.0\textwidth]{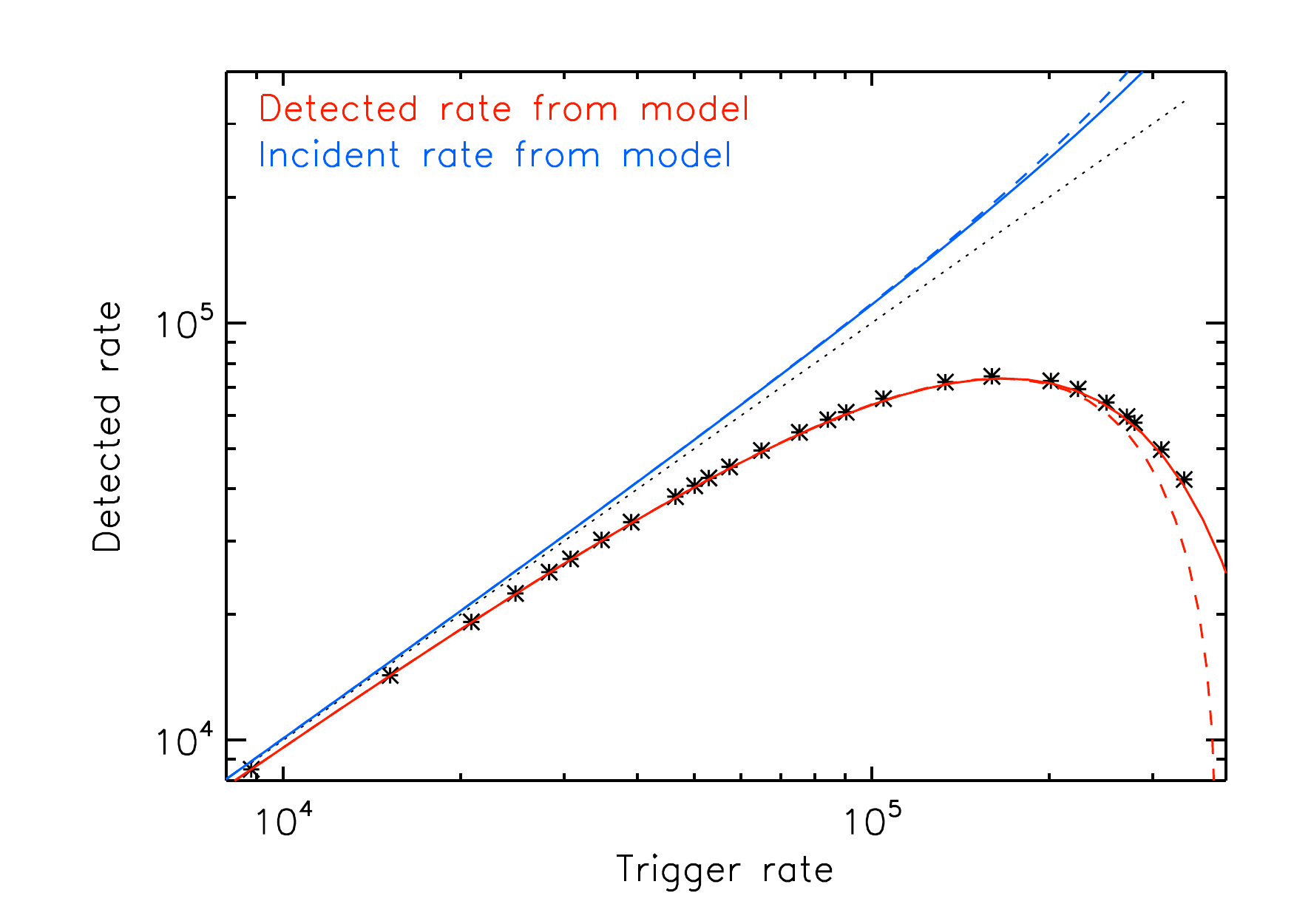}
 \caption{The detected event rate is shown as a function of event trigger rate (black star symbols). 
 Red lines show the expected trends of detected event rate with the trigger rate, and blue lines 
 show the incident rates. Solid and dashed lines correspond to the predictions 
 with \textit{non-paralyzable} and \textit{paralyzable} dead time models for event triggers.}
 \label{deadtime_rates}
 \end{figure}

 In order to verify the dead time behavior of the instrument, the 
data acquired at different incident rates with the miniature X-ray gun, 
as described in section~\ref{crate_sec}, was used.
From the data, trigger rate and the rate of recorded events for 
each case were computed and are plotted as data points in figure~\ref{deadtime_rates}. 

In order to compare with the observations, trigger and detected rates 
for a range of actual incident rates are computed considering the 
non-paralyzable and paralyzable models for trigger and detected 
events, respectively, using the following equations~\cite{knoll00}:

\begin{equation}
n_{t} = \frac{n_{a}}{1+ n_{a} {\tau}_1}
\end{equation}

\begin{equation}
n_{d} = n_{a} ~exp(-n_{a} {\tau}_2)
\end{equation}

\noindent where $n_t, n_d,$ and $n_a$ are the trigger, detected, and actual rates, respectively.
Dead times involved are ${\tau}_1$ and ${\tau}_2$. Model predictions for 
${\tau}_1 = 0.96 ~\mu s$ and ${\tau}_2 = 5 ~\mu s$ are plotted with red solid line
in the figure~\ref{deadtime_rates}, which matches
the observations. 
For comparison, the expected behavior for event triggers in the case of paralyzable model 
is shown with dashed lines that deviate from the observations.  
Blue lines in the figure show the incident rates as a function of trigger rates 
for both cases. 

With the experimental verification of the dead time characteristics of the XSM, 
the model and the obtained dead time values can be used to correct the dead time 
effects in the observed flux. This is done by modifying the exposure time 
of the observed spectrum by:

\begin{equation}    
T_{live} = T_{exp} * \frac{n_d}{n_t} * (1~-~n_t~{\tau}_1)
\end{equation}

\noindent where the symbols have the same meaning as earlier. It can be noted 
that, for low count rates, the correction term is negligible, whereas, at 
higher rates, this becomes important. XSM data analysis software includes this 
correction for the exposure time.   

Another effect that is important at high count rates for X-ray detectors is
the pulse pileup. When the incident rates increase, there is a chance that two 
photons are incident on the detector within the charge-readout time scales, 
which is less than a microsecond in the case of XSM.
In that case, the detector would record only one event with the measured 
energy equivalent to the sum of energies of both photons, skewing the 
measured spectrum to higher energies than the actual. 

 \begin{figure}
 \centerline{\includegraphics[width=0.9\textwidth]{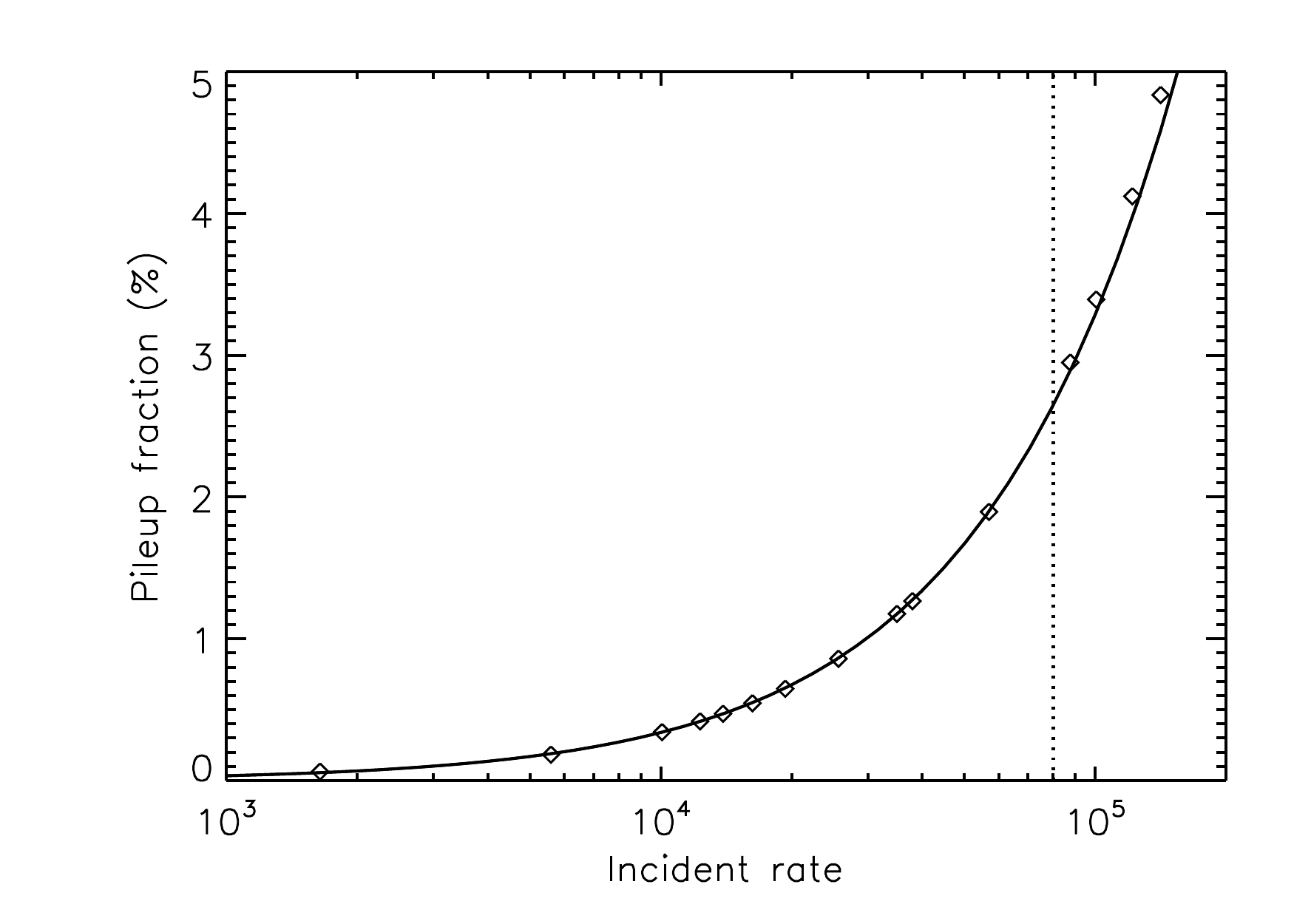}}
 \caption{ Percentage of incident events that are affected by pulse pileup in 
 XSM spectrum as a function of the incident rate as measured from the experiment are 
 shown as the data points. The black solid line is the prediction based on the 
 pulse-peaking time ($0.35~\mu s$) of XSM. The vertical dotted line corresponds to 
 the incident rate of 80,000 $counts~s^{-1}$, the threshold rate for no variation in spectral 
 performance.}
 \label{pileupfrac}
 \end{figure}

In order to investigate the pileup effects in XSM, spectra of the mono-energetic 
line at 5 keV, obtained from the RRCAT facility, with different incident rates
were used. As the incident 
rate increases, the obtained spectrum has events with energies above 5 keV 
that are due to the pileup of photons. The fraction of the total events that are affected 
by pileup was computed by taking the ratio of events in the spectrum above 5 keV 
with the total. This fraction is then corrected for the dead time effects, as 
discussed earlier. Figure~\ref{pileupfrac} shows the measured fraction of incident photons 
that suffered pileup in the XSM detector as a function of incident rate. 
The same obtained from a model with the XSM pulse peaking time of $0.35~\mu s$ 
is overplotted with a solid line and was found to be consistent with 
the measurements.  It may be noted that the pileup fraction is independent 
of incident photon energy; hence, these measurements are directly applicable 
to solar observations. As discussed in the previous section, the XSM is 
expected to operate up to an incident rate of 80,000 $counts~s^{-1}$ until 
which there is no spectral degradation, and it is 
seen from the figure~\ref{pileupfrac} that the pileup fraction is less than 3\% in this 
range. Hence, the effect of pileup can be safely ignored for the spectral analysis.

\section{Summary}
\label{summary}

The results of ground calibration of the Chandrayaan-2 \textit{Solar X-ray Monitor}
are presented. Dedicated experiments were carried out to determine various
parameters of the instrument response. The linearity of the spectrometer
was established over the entire energy range, and the dependency of the gain parameters 
on temperature and interaction position was parameterized from extensive experimental data
such that the uncertainty in energy measurement is less than 10 eV. 
The stability of spectral performance with incident flux was assessed 
and it has been established that it remains unaffected up to a count rate of
80,000 $counts~s^{-1}$ suggesting that spectroscopy with the XSM in 
the 1--15 keV energy range is feasible for flares up to $\sim$M5 
class, and up to $\sim$X5 class with the low energy threshold increased
to 2 keV using the beryllium filter.
For this operating count rate range, the dead time correction method has been 
established, and pileup effects were shown to be unimportant. 
With a suite of observations of mono-energetic X-ray lines, the spectral redistribution function 
of the detector was modeled with empirical functions having energy dependant 
parameters, taking into account all features to provide an accurate redistribution matrix.
The on-ground estimate of the effective area as a function of energy and angle was 
obtained from the experimental measurements of the geometric area and collimator response, 
and the manufacturer provided values of the detector module parameters.

Thus, all aspects of the XSM response, required to infer the incident photon 
spectrum from the observed count spectrum, have been well-calibrated. 
The Chandrayaan-2 mission was launched on 22 July 2019, and the 
XSM began its operations in early September 2019. 
Preliminary observations suggest that the instrument is performing as 
expected~\footnote{See the update dated October 10, 2019 at \url{https://www.isro.gov.in/chandrayaan2-latest-updates}}. 
Details of in-flight performance and refinement of the calibration will be reported elsewhere.

%

\begin{acknowledgements}
The XSM payload was designed and developed by Physical Research Laboratory (PRL), Ahmedabad, 
supported by the Department of Space, Govt. of India. 
PRL was also responsible for the development of the required
data processing software, overall payload operations, and data analysis
of XSM. The filter wheel mechanism for the XSM was provided by U. R. Rao
Satellite Centre (URSC), Bengaluru, along with Laboratory for Electro-Optics 
Systems (LEOS), Bengaluru. Thermal design and analysis of
the XSM packages were carried out by URSC whereas Space Application
Centre (SAC) supported in mechanical design and analysis. SAC
also supported in the fabrication of the flight model of the payload and
its test and evaluation for the flight use. We thank various
facilities and the technical teams of all the above centres for their support
during the design, fabrication, and testing of this payload. 
X-ray beam lines BL-03 and BL-16 of the Indus-2 synchrotron facility at the Raja Ramanna Center for 
Advanced Technology, Indore were used in the calibration of the XSM. 
The Chandrayaan-2 project is funded and managed by the Indian Space Research Organisation (ISRO).
\end{acknowledgements}

%


\bibliographystyle{spphys}
\bibliography{references_xsm.bib}  

\begin{thebibliography}{10}
\providecommand{\url}[1]{{#1}}
\providecommand{\urlprefix}{URL }
\expandafter\ifx\csname urlstyle\endcsname\relax
  \providecommand{\doi}[1]{DOI \discretionary{}{}{}#1}\else
  \providecommand{\doi}{DOI \discretionary{}{}{}\begingroup
  \urlstyle{rm}\Url}\fi

\bibitem{vanitha20}
M.~{Vanitha}, P.~{Veeramuthuvel}, K.~{Kalpana}, G.~{Nagesh}, in \emph{Lunar and
  Planetary Science Conference} (2020), Lunar and Planetary Science Conference,
  p. 1994

\bibitem{radhakrishna20}
V.~{Radhakrishna}, A.~{Tyagi}, S.~{Narendranath}, K.~{Vadodariya}, R.~{Yadav},
  B.~{Singh}, G.~{Balaji}, N.~{Satya}, A.~{Shetty}, H.N. {Suresh Kumar},
  {Kumar}, S.~{Vaishali}, N.S. {Pillai}, S.~{Tadepalli}, V.~{Raghavendra},
  P.~{Sreekumar}, A.~{Agarwal}, N.~{Valarmathi}, Current Science
  \textbf{118}(2), 219 (2020).
\newblock \doi{10.18520/cs/v118/i2/219-225}

\bibitem{shanmugam20}
M.~{Shanmugam}, S.V. {Vadawale}, A.R. {Patel}, H.K. {Adalaja}, N.P.S. {Mithun},
  T.~{Ladiya}, S.K. {Goyal}, N.K. {Tiwari}, N.~{Singh}, S.~{Kumar}, D.K.
  {Painkra}, Y.B. {Acharya}, A.~{Bhardwaj}, A.K. {Hait}, A.~{Patinge}, A.h.
  {Kapoor}, H.N.S. {Kumar}, N.~{Satya}, G.~{Saxena}, K.~{Arvind}, Current
  Science \textbf{118}(1), 45 (2020).
\newblock \doi{10.18520/cs/v118/i1/45-52}

\bibitem{2015A&A...582A...4J}
B.~{Joshi}, R.~{Bhattacharyya}, K.K. {Pandey}, U.~{Kushwaha}, Y.J. {Moon}, \aap
  \textbf{582}, A4 (2015).
\newblock \doi{10.1051/0004-6361/201526369}

\bibitem{modi19}
M.~Modi, R.~Gupta, S.~Kane, V.~Prasad, C.~Garg, P.~Yadav, V.~Raghuvanshi,
  A.~Singh, M.~Sinha, in \emph{AIP Conference Proceedings}, vol. 2054 (2019),
  vol. 2054, p. 060022.
\newblock \doi{10.1063/1.5084653}

\bibitem{tiwari13}
M.~Tiwari, P.~Gupta, A.~Sinha, S.~Kane, A.~Singh, S.~Garg, C.~Garg, G.~Lodha,
  S.~Deb, Journal of synchrotron radiation \textbf{20}, 386 (2013).
\newblock \doi{10.1107/S0909049513001337}

\bibitem{scholze09}
F.~{Scholze}, M.~{Procop}, X-ray Spectrometry \textbf{38}(4), 312 (2009).
\newblock \doi{10.1002/xrs.1165}

\bibitem{arnaud96}
K.A. {Arnaud}, in \emph{Astronomical Data Analysis Software and Systems V},
  \emph{Astronomical Society of the Pacific Conference Series}, vol. 101, ed.
  by G.H. {Jacoby}, J.~{Barnes} (1996), \emph{Astronomical Society of the
  Pacific Conference Series}, vol. 101, p.~17

\bibitem{agostinelli03}
S.~Agostinelli, J.~Allison, K.~Amako, J.~Apostolakis, H.~Araujo, P.~Arce,
  M.~Asai, D.~Axen, S.~Banerjee, G.~Barrand, F.~Behner, L.~Bellagamba,
  J.~Boudreau, L.~Broglia, A.~Brunengo, H.~Burkhardt, S.~Chauvie, J.~Chuma,
  R.~Chytracek, G.~Cooperman, G.~Cosmo, P.~Degtyarenko, A.~Dell'Acqua,
  G.~Depaola, D.~Dietrich, R.~Enami, A.~Feliciello, C.~Ferguson, H.~Fesefeldt,
  G.~Folger, F.~Foppiano, A.~Forti, S.~Garelli, S.~Giani, R.~Giannitrapani,
  D.~Gibin, J.G. Cadenas, I.~González, G.G. Abril, G.~Greeniaus, W.~Greiner,
  V.~Grichine, A.~Grossheim, S.~Guatelli, P.~Gumplinger, R.~Hamatsu,
  K.~Hashimoto, H.~Hasui, A.~Heikkinen, A.~Howard, V.~Ivanchenko, A.~Johnson,
  F.~Jones, J.~Kallenbach, N.~Kanaya, M.~Kawabata, Y.~Kawabata, M.~Kawaguti,
  S.~Kelner, P.~Kent, A.~Kimura, T.~Kodama, R.~Kokoulin, M.~Kossov,
  H.~Kurashige, E.~Lamanna, T.~Lampén, V.~Lara, V.~Lefebure, F.~Lei,
  M.~Liendl, W.~Lockman, F.~Longo, S.~Magni, M.~Maire, E.~Medernach,
  K.~Minamimoto, P.M. de~Freitas, Y.~Morita, K.~Murakami, M.~Nagamatu,
  R.~Nartallo, P.~Nieminen, T.~Nishimura, K.~Ohtsubo, M.~Okamura, S.~O'Neale,
  Y.~Oohata, K.~Paech, J.~Perl, A.~Pfeiffer, M.~Pia, F.~Ranjard, A.~Rybin,
  S.~Sadilov, E.D. Salvo, G.~Santin, T.~Sasaki, N.~Savvas, Y.~Sawada,
  S.~Scherer, S.~Sei, V.~Sirotenko, D.~Smith, N.~Starkov, H.~Stoecker,
  J.~Sulkimo, M.~Takahata, S.~Tanaka, E.~Tcherniaev, E.S. Tehrani, M.~Tropeano,
  P.~Truscott, H.~Uno, L.~Urban, P.~Urban, M.~Verderi, A.~Walkden, W.~Wander,
  H.~Weber, J.~Wellisch, T.~Wenaus, D.~Williams, D.~Wright, T.~Yamada,
  H.~Yoshida, D.~Zschiesche, Nuclear Instruments and Methods in Physics
  Research Section A: Accelerators, Spectrometers, Detectors and Associated
  Equipment \textbf{506}(3), 250  (2003).
\newblock \doi{https://doi.org/10.1016/S0168-9002(03)01368-8}.
\newblock
  \urlprefix\url{http://www.sciencedirect.com/science/article/pii/S0168900203013688}

\bibitem{1997A&AS..125..149D}
K.P. {Dere}, E.~{Landi}, H.E. {Mason}, B.C. {Monsignori Fossi}, P.R. {Young},
  \aaps \textbf{125}, 149 (1997).
\newblock \doi{10.1051/aas:1997368}

\bibitem{2015A&A...582A..56D}
G.~{Del Zanna}, K.P. {Dere}, P.R. {Young}, E.~{Landi}, H.E. {Mason}, \aap
  \textbf{582}, A56 (2015).
\newblock \doi{10.1051/0004-6361/201526827}

\bibitem{knoll00}
G.F. {Knoll}, \emph{{Radiation detection and measurement}} (2000)

\end{thebibliography}


\end{document}